\documentclass[aps,prd,twocolumn,floats,floatfix,nofootinbib]{revtex4-1}
\usepackage{graphicx}
\usepackage{amssymb}
\usepackage{epstopdf}
\usepackage{hyperref}

\usepackage{amsmath}
\usepackage{bbold}
\usepackage{color}

\begin{document}

\title{Real singlet scalar dark matter extension of the Georgi-Machacek model}

\author{Robyn Campbell}
\email{rcampbel@physics.carleton.ca}

\author{Stephen Godfrey}
\email{godfrey@physics.carleton.ca}

\author{Heather E.~Logan}
\email{logan@physics.carleton.ca} 

\author{Alexandre Poulin}
\email{apoulin@physics.carleton.ca}

\affiliation{Ottawa-Carleton Institute for Physics, Carleton University, 1125 Colonel By Drive, Ottawa, Ontario K1S 5B6, Canada}

\date{October 26, 2016}                                  

\begin{abstract}
The Georgi-Machacek model extends the Standard Model Higgs sector with the addition of isospin-triplet scalar fields 
in such a way as to preserve the custodial symmetry.  The presence of higher-isospin scalars contributing to 
electroweak symmetry breaking offers the interesting possibility that the couplings of the 125~GeV Higgs boson 
to both gluons and vector boson pairs could be larger than those of the Standard Model Higgs boson. 
 Constraining this possibility using measurements of Higgs production and decay at the CERN Large 
 Hadron Collider is notoriously problematic if a new, non-Standard Model decay mode of the 125~GeV Higgs 
 boson is present.  We study an implementation of this scenario in which the Georgi-Machacek model is 
 extended by a real singlet scalar dark matter candidate, and require that the singlet scalar account 
 for all the dark matter in the universe.  The combination of the observed dark matter relic density and 
 direct detection constraints exclude singlet scalar masses below about 57~GeV. 
 Higgs measurements are not yet precise enough to be very sensitive 
 to $h\rightarrow SS$ in the remaining allowed kinematic region,
 so that constraints from Higgs measurements 
are so far the same as in the GM model without a singlet scalar.
We also find that, above the Higgs pole, a substantial region of parameter space yielding 
the correct dark matter relic density can escape the near-future direct detection experiments 
DEAP and XENON~1T for dark matter masses as low as 120~GeV and even have a direct detection cross 
section below the neutrino floor for $m_S\gtrsim 150$~GeV.  This is in contrast to the singlet scalar 
dark matter extension of the Standard Model, for which these future experiments are expected to 
exclude dark matter masses above the Higgs pole up to the multi-TeV range.
\end{abstract}

\maketitle 

\section{Introduction}

Since the discovery of a Standard Model (SM)-like 125~GeV Higgs boson at the CERN Large Hadron Collider (LHC)~\cite{Aad:2012tfa}, the determination of the Higgs boson's couplings to other particles has become a top priority.  At the LHC, these couplings are extracted from signal rates in various resonant Higgs production and decay channels, which can be written in the narrow width approximation as
\begin{equation}
	{\rm Rate}_{ij} = \sigma_i \frac{\Gamma_j}{\Gamma_{\rm tot}}
	= \kappa_i^2 \sigma_i^{\rm SM} \frac{\kappa_j^2 \Gamma_j^{\rm SM}}
	{\sum_k \kappa_k^2 \Gamma_k^{\rm SM} + \Gamma_{\rm new}}.
\end{equation}
Here $\sigma_i$ is the Higgs production cross section in production mode $i$, $\Gamma_j$ is the Higgs decay partial width into final state $j$, $\Gamma_{\rm tot}$ is the total width of the Higgs boson, the corresponding quantities in the SM are denoted with a superscript, and $\Gamma_{\rm new}$ represents the partial width of the Higgs boson into any new, non-SM final states.  The coupling modification factors $\kappa_i$ parameterize the deviations of the Higgs couplings from their SM values~\cite{LHCHiggsCrossSectionWorkingGroup:2012nn}.

The extraction of the Higgs couplings $\kappa_i$ from these LHC rate measurements is plagued by a well-known ``flat direction''~\cite{Zeppenfeld:2000td} that appears when new decay modes are present.  
For example, one can imagine a scenario in which all the coupling modification factors have a common value $\kappa_i^2 \equiv \kappa^2 > 1$ and there is a new, unobserved contribution to the Higgs total width, $\Gamma_{\rm new} > 0$.  In this case the Higgs production and decay rates measurable at the LHC are given by
\begin{equation}
	{\rm Rate}_{ij} = \frac{\kappa^4 \sigma_i^{\rm SM} \Gamma_j^{\rm SM}}{\kappa^2 \Gamma_{\rm tot}^{\rm SM} + \Gamma_{\rm new}}.
\end{equation}
All measured Higgs production and decay rates will be equal to their SM values if
\begin{equation}
	\Gamma_{\rm new} = \kappa^2 (\kappa^2 - 1) \Gamma_{\rm tot}^{\rm SM} \geq 0.
\end{equation}
In particular, a simultaneous enhancement $\kappa^2 > 1$ of all the Higgs couplings to SM particles can mask, and be masked by, the presence of new decay modes of the Higgs that are not (yet) directly detected at the LHC.\footnote{Measuring such an enhancement in the Higgs couplings would be straightforward at a lepton-collider Higgs factory such as the International Linear Collider (ILC), where a direct measurement of the total Higgs production cross section in $e^+e^- \to Zh$ can be made with no reference to the Higgs decay branching ratios by using the recoil mass method (see, e.g., Ref.~\cite{Baer:2013cma}).}

Our goal in this paper is to study an explicit benchmark model in which this scenario could be realized.  
We focus on models with extended Higgs sectors.  Our first requirement is a model in which the Higgs 
couplings to $W$ and $Z$ bosons and to fermions can be enhanced relative to those in the SM.  
To achieve $\kappa_W, \kappa_Z > 1$ in an extended Higgs model, we need scalars in isospin 
representations larger than doublets that carry non-negligible vacuum expectation values (vevs).  
Only a few such models exist that preserve the $\rho$ parameter at tree level: the Georgi-Machacek 
(GM) model with isospin triplets~\cite{Georgi:1985nv,Chanowitz:1985ug}, generalizations of the 
GM model to higher isospin~\cite{Galison:1983qg,Robinett:1985ec,Logan:1999if,Chang:2012gn,Logan:2015xpa}, 
and an extension of the Higgs sector by an isospin septet with appropriately-chosen 
hypercharge~\cite{Hisano:2013sn,Kanemura:2013mc,Alvarado:2014jva}.  In this paper we 
choose the GM model as the simplest extension suitable for our purposes.  Its phenomenology 
has been extensively 
studied~\cite{Gunion:1989ci,Gunion:1990dt,HHG,Haber:1999zh,Aoki:2007ah,Godfrey:2010qb,Low:2010jp,Logan:2010en,Falkowski:2012vh,Chang:2012gn,Chiang:2012cn,Chiang:2013rua,Kanemura:2013mc,Englert:2013zpa,Killick:2013mya,Englert:2013wga,Efrati:2014uta,Hartling:2014zca,Chiang:2014hia,Chiang:2014bia,Godunov:2014waa,Hartling:2014aga,Chiang:2015kka,Godunov:2015lea}.  
It has also been incorporated into the scalar sectors of little Higgs~\cite{Chang:2003un,Chang:2003zn} 
and supersymmetric~\cite{Cort:2013foa,Garcia-Pepin:2014yfa} models, and an extension with an 
additional isospin doublet~\cite{Hedri:2013wea} has been considered.

Our second requirement is a new decay mode for the 125~GeV Higgs boson.  A particularly attractive prospect is to link Higgs physics to the mystery of dark matter in the universe (for a recent pedagogical review see Ref.~\cite{Gelmini:2015zpa}) by allowing the Higgs to decay into pairs of dark matter particles.  To this end we extend the GM model through the addition of a real isospin-singlet scalar field $S$, upon which we impose a $Z_2$ symmetry $S \to -S$.  We will require that $S$ accounts for the observed dark matter relic abundance in the universe via the standard thermal freeze-out mechanism.  Real singlet scalar extensions of the SM~\cite{Veltman:1989vw,Silveira:1985rk,McDonald:1993ex,Burgess:2000yq,McDonald:2001vt,Barger:2007im,Goudelis:2009zz,Gonderinger:2009jp,He:2009yd,Profumo:2010kp,Yaguna:2011qn,Drozd:2011aa,Djouadi:2011aa,Kadastik:2011aa,Djouadi:2012zc,Cheung:2012xb,Damgaard:2013kva,Cline:2013gha,Baek:2014jga,Feng:2014vea,Campbell:2015fra} and of two-Higgs-doublet models~\cite{He:2008qm,Grzadkowski:2009iz,Logan:2010nw,Boucenna:2011hy,He:2011gc,Bai:2012nv,He:2013suk,Cai:2013zga,Wang:2014elb,Chen:2013jvg,Drozd:2014yla,Wang:2014elb,Campbell:2015fra} have been extensively studied in the literature.  These models tend to be tightly constrained by the combination of relic density, dark matter direct-detection limits, and limits on the indirect detection of dark matter annihilation byproducts from nearby dwarf galaxies.

We will find that the situation is rather similar in the singlet scalar dark matter extension of the GM model.  
The two strongest constraints are the requirement of the correct 
dark matter relic abundance from thermal freeze-out \cite{Kolb:1990vq}
and the direct detection cross section limit from the LUX experiment~\cite{Akerib:2016vxi}.  
These constraints restrict the allowed range of singlet scalar masses to lie just below the 125~GeV Higgs pole 
for resonant annihilation (57--62~GeV) or above the $Z$ boson mass.  The constraint from 125~GeV Higgs 
boson invisible decays is currently weaker than that from direct detection.  Constraints coming from Higgs 
properties and signals also significantly constrain this model. They do however allow for some interesting 
deviations from the Standard Model that the GM model without the singlet does not allow.

One important difference compared to the singlet scalar extension of the SM is the prospect for 
future dark matter direct detection experiments to probe the model at heavier singlet masses.  
While an absence of signal at the planned XENON~1T experiment would exclude singlet scalar masses 
up to 4.5~TeV in the singlet scalar extension of the SM~\cite{Cline:2013gha}, in the singlet 
scalar extension of the GM model a large swath of parameter space with singlet scalar masses as 
light as 125~GeV remains beyond the reach of XENON~1T. In fact, there is some allowed parameter 
space with singlet scalar masses near the 125 GeV Higgs pole for resonant annihilation (60--62~GeV) 
and some with singlet scalar masses above about 150~GeV which have a direct detection cross section 
that lies below the neutrino floor.  This is mainly due to the contribution of the additional 
scalars in the GM model to the production of the correct relic density, while not contributing 
strongly to the direct detection cross section. 

This paper is organized as follows.  In Sec.~\ref{sec:model} we begin with a description of the singlet 
scalar extension of the GM model.  In Sec.~\ref{sec:theory_constraints} we extend the theoretical 
constraints on the GM model to include the singlet scalar extension.  
In Sec.~\ref{sec:freezeout} we describe the details of the thermal freezeout and imposing 
the relic abundance constraints on the model parameters while in Sec.~\ref{sec:Scan} we
describe the numerical scan procedure used to map out the allowed parameter space.
In Sec.~\ref{sec:collider}
 we briefly summarize the direct and indirect search constraints on the additional scalars in the 
 GM model.  In Sec.~\ref{sec:darkmatter} we compute the dark matter relic abundance and direct and 
 indirect detection cross sections and display the impact of the observational constraints on the 
 allowed parameter space.  In Sec.~\ref{sec:consequences} we consider the constraints from the 
 125~GeV Higgs boson invisible decays and signal strengths in visible channels.  Finally 
 in Sec.~\ref{sec:conclusions} we summarize our conclusions.  Feynman rules for couplings involving the 
 singlet scalar are collected in an appendix.

\section{The Georgi-Machacek model extended by a real singlet scalar}
\label{sec:model}

The scalar sector of the GM model~\cite{Georgi:1985nv,Chanowitz:1985ug} consists of the usual complex doublet $(\phi^+,\phi^0)$ with hypercharge\footnote{We use $Q = T^3 + Y/2$.} $Y = 1$, a real triplet $(\xi^+,\xi^0,\xi^-)$ with $Y = 0$, and  a complex triplet $(\chi^{++},\chi^+,\chi^0)$ with $Y=2$.  The doublet is responsible for the fermion masses as in the SM.
In order to preserve the custodial SU(2) symmetry and avoid large tree-level contributions to the electroweak $\rho$ parameter, the scalar potential is constructed to preserve a global SU(2)$_L \times$SU(2)$_R$ symmetry, which breaks down to the diagonal subgroup (known as the custodial SU(2) symmetry) upon electroweak symmetry breaking.  To make the global SU(2)$_L \times$SU(2)$_R$ symmetry explicit, we write the doublet in the form of a bidoublet $\Phi$ and combine the triplets to form a bitriplet $X$:
\begin{equation}
	\Phi = \left( \begin{array}{cc}
	\phi^{0*} &\phi^+  \\
	-\phi^{+*} & \phi^0  \end{array} \right), \qquad
	X =
	\left(
	\begin{array}{ccc}
	\chi^{0*} & \xi^+ & \chi^{++} \\
	 -\chi^{+*} & \xi^{0} & \chi^+ \\
	 \chi^{++*} & -\xi^{+*} & \chi^0  
	\end{array}
	\right).
	\label{eq:PX}
\end{equation}
The vacuum expectation values (vevs) are defined by 
$\langle \Phi  \rangle = \frac{ v_{\phi}}{\sqrt{2}} I_{2\times2}$  
and $\langle X \rangle = v_{\chi} I_{3 \times 3}$, where $I$ is the unit matrix and 
the Fermi constant $G_F$ constrains
\begin{equation}
	v_{\phi}^2 + 8 v_{\chi}^2 \equiv v^2 = \frac{1}{\sqrt{2} G_F} \approx (246~{\rm GeV})^2.
	\label{eq:vevrelation}
\end{equation} 

The most general gauge-invariant scalar potential involving these fields and the real singlet $S$, 
while conserving the global SU(2)$_L\times$SU(2)$_R$ and the $Z_2$ symmetry $S \to -S$, is given by
\begin{widetext}
\begin{eqnarray}
	V(\Phi,X) &= & \frac{\mu_2^2}{2}  \text{Tr}(\Phi^\dagger \Phi) 
	+  \frac{\mu_3^2}{2}  \text{Tr}(X^\dagger X)  
	+ \lambda_1 [\text{Tr}(\Phi^\dagger \Phi)]^2  
	+ \lambda_2 \text{Tr}(\Phi^\dagger \Phi) \text{Tr}(X^\dagger X)   \nonumber \\
          & & + \lambda_3 \text{Tr}(X^\dagger X X^\dagger X)  
          + \lambda_4 [\text{Tr}(X^\dagger X)]^2 
           - \lambda_5 \text{Tr}( \Phi^\dagger \tau^a \Phi \tau^b) \text{Tr}( X^\dagger t^a X t^b) 
           \nonumber \\
           & & - M_1 \text{Tr}(\Phi^\dagger \tau^a \Phi \tau^b)(U X U^\dagger)_{ab}  
           -  M_2 \text{Tr}(X^\dagger t^a X t^b)(U X U^\dagger)_{ab} \nonumber \\
           & & + \frac{\mu_S^2}{2} S^2 + \lambda_a S^2 {\rm Tr}(\Phi^{\dagger} \Phi)
	  + \lambda_b S^2 {\rm Tr}(X^{\dagger} X) + \lambda_S S^4.
            \label{eq:potential}
\end{eqnarray} 
\end{widetext}
The first three lines of this potential are identical to that given, e.g., in Ref.~\cite{Hartling:2014zca}.\footnote{A translation table to other parameterizations of the GM model scalar potential has been given in the appendix of Ref.~\cite{Hartling:2014zca}.}  The last line contains the new terms involving the singlet scalar $S$.
Here the SU(2)$_L$ generators for the doublet representation are $\tau^a = \sigma^a/2$ with $\sigma^a$ being the Pauli matrices, the generators for the triplet representation are
\begin{eqnarray}
	t^1&=& \frac{1}{\sqrt{2}} \left( \begin{array}{ccc}
	 0 & 1  & 0  \\
	  1 & 0  & 1  \\
	  0 & 1  & 0 \end{array} \right), \qquad  
	  t^2= \frac{1}{\sqrt{2}} \left( \begin{array}{ccc}
	 0 & -i  & 0  \\
	  i & 0  & -i  \\
	  0 & i  & 0 \end{array} \right), \nonumber \\
	t^3 &=& \left( \begin{array}{ccc}
	 1 & 0  & 0  \\
	  0 & 0  & 0  \\
	  0 & 0 & -1 \end{array} \right),
\end{eqnarray}
and the matrix $U$, which rotates $X$ into the Cartesian basis, is given by~\cite{Aoki:2007ah}
\begin{equation}
	 U = \left( \begin{array}{ccc}
	- \frac{1}{\sqrt{2}} & 0 &  \frac{1}{\sqrt{2}} \\
	 - \frac{i}{\sqrt{2}} & 0  &   - \frac{i}{\sqrt{2}} \\
	   0 & 1 & 0 \end{array} \right).
	 \label{eq:U}
\end{equation}

We will work in the vacuum in which $S$ does not get a vev, so that the $Z_2$ symmetry remains 
unbroken and $S$ is stable.  The presence of $S$ then has no effect on the mass spectrum or 
potential-minimization conditions of the GM sector of the model, which can be taken from 
Ref.~\cite{Hartling:2014zca}.  We summarize the physical spectrum here.

The physical fields can be organized by their transformation properties under the custodial SU(2) symmetry into a custodial fiveplet, a custodial triplet, and three custodial singlets, one of which is $S$.  The custodial-fiveplet and -triplet states are given by
\begin{eqnarray}
	&& H_5^{++} = \chi^{++}, \qquad
	H_5^+ = \frac{\left(\chi^+ - \xi^+\right)}{\sqrt{2}},   \nonumber \\
&&	H_5^0 = \sqrt{\frac{2}{3}} \xi^0 - \sqrt{\frac{1}{3}} \chi^{0,r}, 	\nonumber \\
	&& H_3^+ = - s_H \phi^+ + c_H \frac{\left(\chi^++\xi^+\right)}{\sqrt{2}},  \nonumber \\
	&& H_3^0 = - s_H \phi^{0,i} + c_H \chi^{0,i},
\end{eqnarray}
and their complex conjugates,
where the vevs are parameterized by
\begin{equation}
	c_H \equiv \cos\theta_H = \frac{v_{\phi}}{v}, \qquad
	s_H \equiv \sin\theta_H = \frac{2\sqrt{2}\,v_\chi}{v}, \label{eq:cHsH}
\end{equation}
and we have decomposed the neutral fields into real and imaginary parts according to
\begin{eqnarray}
&&	\phi^0 \to \frac{v_{\phi}}{\sqrt{2}} + \frac{\phi^{0,r} + i \phi^{0,i}}{\sqrt{2}},
	\qquad
	\chi^0 \to v_{\chi} + \frac{\chi^{0,r} + i \chi^{0,i}}{\sqrt{2}}, \nonumber \\
&&	\xi^0 \to v_{\chi} + \xi^0.
	\label{eq:decomposition}
\end{eqnarray}
The masses within each custodial multiplet are degenerate at tree 
level and can be written (after eliminating $\mu_2^2$ and $\mu_3^2$ in favor of the vevs) 
as\footnote{Note that the ratio $M_1/v_{\chi}$ is finite in the limit $v_{\chi} \to 0$, 
\begin{equation}
	\frac{M_1}{v_{\chi}} = \frac{4}{v_{\phi}^2} 
	\left[ \mu_3^2 + (2 \lambda_2 - \lambda_5) v_{\phi}^2 
	+ 4(\lambda_3 + 3 \lambda_4) v_{\chi}^2 - 6 M_2 v_{\chi} \right], 
\end{equation}
which follows from the minimization condition $\partial V/\partial v_{\chi} = 0$~\cite{Hartling:2014zca}.}
\begin{eqnarray}
	m_5^2 &=& \frac{M_1}{4 v_{\chi}} v_\phi^2 + 12 M_2 v_{\chi} 
	+ \frac{3}{2} \lambda_5 v_{\phi}^2 + 8 \lambda_3 v_{\chi}^2,  \\
	m_3^2 &=&  \frac{M_1}{4 v_{\chi}} (v_\phi^2 + 8 v_{\chi}^2) 
	+ \frac{\lambda_5}{2} (v_{\phi}^2 + 8 v_{\chi}^2) 
	= \left(  \frac{M_1}{4 v_{\chi}} + \frac{\lambda_5}{2} \right) v^2.  \nonumber
\end{eqnarray}

The gauge singlet $S$ remains a mass eigenstate, with physical mass-squared given by 
\begin{equation}
	m_S^2 = \mu_S^2 + 2 \lambda_a v_{\phi}^2 + 6 \lambda_b v_{\chi}^2,
	\label{eq:m_s}
\end{equation}
which we require to be positive to avoid breaking the $Z_2$ symmetry.

The other two custodial SU(2)--singlet mass eigenstates are given by
\begin{eqnarray}
	h &=& \cos \alpha \, \phi^{0,r} - \sin \alpha \, H_1^{0\prime},  \nonumber \\
	H &=& \sin \alpha \, \phi^{0,r} + \cos \alpha \, H_1^{0\prime},
	\label{mh-mH}
\end{eqnarray}
where 
\begin{equation}
	H_1^{0 \prime} = \sqrt{\frac{1}{3}} \xi^0 + \sqrt{\frac{2}{3}} \chi^{0,r}.
\end{equation}
The mixing angle and masses are given by
\begin{eqnarray}
	&&\sin 2 \alpha =  \frac{2 \mathcal{M}^2_{12}}{m_H^2 - m_h^2},    \qquad
	\cos 2 \alpha =  \frac{ \mathcal{M}^2_{22} - \mathcal{M}^2_{11}  }{m_H^2 - m_h^2}, 
	 \\
	&&m^2_{h,H} = \frac{1}{2} \left[ \mathcal{M}_{11}^2 + \mathcal{M}_{22}^2
	\mp \sqrt{\left( \mathcal{M}_{11}^2 - \mathcal{M}_{22}^2 \right)^2 
	+ 4 \left( \mathcal{M}_{12}^2 \right)^2} \right],  \nonumber
	\label{eq:hmass}
\end{eqnarray}
where we choose $m_h < m_H$, and 
\begin{eqnarray}
	\mathcal{M}_{11}^2 &=& 8 \lambda_1 v_{\phi}^2, \nonumber \\
	\mathcal{M}_{12}^2 &=& \frac{\sqrt{3}}{2} v_{\phi} 
	\left[ - M_1 + 4 \left(2 \lambda_2 - \lambda_5 \right) v_{\chi} \right], \nonumber \\
	\mathcal{M}_{22}^2 &=& \frac{M_1 v_{\phi}^2}{4 v_{\chi}} - 6 M_2 v_{\chi} 
	+ 8 \left( \lambda_3 + 3 \lambda_4 \right) v_{\chi}^2.
\end{eqnarray}

\section{Theoretical Constraints on Lagrangian Parameters}
\label{sec:theory_constraints}

The singlet scalar dark matter extension of the GM model has 13 free parameters, two of which can be fixed by $G_F$ and the 125~GeV Higgs mass.  Before scanning over the remaining parameters, we first study the relevant theoretical and experimental constraints.  The theoretical constraints come from (1) perturbative unitarity imposed on $2\to 2$ scalar scattering amplitudes,
(2) the requirement that the scalar potential be bounded from below, and (3) that the custodial SU(2)-preserving minimum is the true global minimum of the potential.

\subsection{Perturbative unitarity of $2 \to 2$ scattering amplitudes}

The scalar couplings in Eq.~\ref{eq:potential} can be bounded by perturbative unitarity 
of the 2 $\rightarrow$ 2 scalar field scattering amplitudes.  These constraints were studied in the original GM model in Refs.~\cite{Aoki:2007ah,Hartling:2014zca}; here we extend them to include the real singlet scalar.

The partial wave amplitudes $a_J$ are related 
to the matrix element $\mathcal{M}$ of the process by:
\begin{equation}
	\mathcal{M}=16\pi\sum_J (2J+1)a_J P_J(\cos\theta),
\end{equation}
where $J$ is the (orbital) angular momentum and $P_J(\cos \theta)$ are the Legendre polynomials. 
Perturbative unitarity requires that the zeroth partial wave amplitude, $a_0$, satisfy $|a_0| \leq 1$ or $|{\rm Re} \, a_0| \leq \frac{1}{2}$. 
Because the 2 $\rightarrow$ 2 scalar field scattering amplitudes are real at tree level, 
we adopt the second, more stringent, constraint. 
We will use this to constrain the magnitudes of the scalar quartic couplings $\lambda_i$. 

We work in the high energy limit, in which the only tree-level diagrams that contribute 
to  $2\rightarrow 2$ scalar scattering are those involving the four-point scalar couplings since 
all diagrams involving scalar propagators are suppressed by the square of the collision energy. 
Thus the dimensionful couplings $M_1$, $M_2$, $\mu_2^2$, $\mu_3^2$ and $\mu_S^2$ are not 
constrained directly by perturbative unitarity. In the high energy limit we can ignore 
electroweak symmetry breaking and include the Goldstone bosons as physical fields 
(this is equivalent to including scattering processes involving longitudinally polarized
 $W$ and $Z$ bosons). We neglect scattering processes involving transversely polarized gauge 
 bosons or fermions. 

Under these conditions, only the zeroth partial wave amplitude contributes to $\mathcal{M}$, so that 
the constraint $|{\rm Re} \, a_0| < \frac{1}{2}$ corresponds to $|\mathcal{M}| < 8\pi$. 
This condition must be applied to each of the eigenvalues of the coupled-channel scattering 
matrix $\mathcal{M}$ including each possible combination of two scalar fields in the initial and 
final states. Because the scalar potential is invariant under $\mbox{SU}(2)_L \times \mbox{U}(1)_Y$, 
the scattering processes preserve electric charge and hypercharge and can be conveniently 
classified by the total electric charge and hypercharge of the incoming and outgoing states. 
We include a symmetry factor of $1/\sqrt{2}$ for each pair of identical particles in the 
initial and final states. 

The basis states for $Q=Y=0$ are,
\begin{eqnarray}
&&	\chi^{++*}\chi^{++}, \ \chi^{+*}\chi^{+}, \ \xi^{+*}\xi^{+}, \ \phi^{+*}\phi^{+}, \ \nonumber \\
&&	\frac{\xi^{0}\xi^{0}}{\sqrt{2}}, \ \chi^{0*}\chi^{0}, \ \phi^{0*}\phi^{0}, \ 
	\frac{S^2}{\sqrt{2}}, \ S\xi^0.
\end{eqnarray}
Scattering amplitudes involving these states yield eight distinct eigenvalues of $\mathcal{M}$,
\begin{align}
x_2^{\pm}&=4\lambda_1-2\lambda_3+4\lambda_4\pm\sqrt{(4\lambda_1+2\lambda_3-4\lambda_4)^2+4\lambda_5^2}, \nonumber \\
y_1&=16\lambda_3+8\lambda_4, \nonumber \\
y_2&=4\lambda_3+8\lambda_4, \nonumber \\
z_b&=4\lambda_b, \nonumber \\
z_{1,2,3} &= {\rm Roots}(P(z)),
\label{eq:eval1}
\end{align}
where $z_1$, $z_2$, and $z_3$ are the roots of the polynomial,
\begin{equation}
P(z)=\det\begin{pmatrix}
24\lambda_1-z 	& 12\lambda_2 					& 4\lambda_a \\
12\lambda_2 	& 28\lambda_3 +44\lambda_4-z	& 6\lambda_b \\
4\lambda_a		& 6\lambda_b					& 12\lambda_S-z
\end{pmatrix}. \label{ref:p}
\end{equation}
We have followed the notation of Refs.~\cite{Aoki:2007ah,Hartling:2014zca} where possible. Note that the pair of eigenvalues $x_1^\pm$ of Refs.~\cite{Aoki:2007ah,Hartling:2014zca} is recovered by taking $\lambda_a=\lambda_b=\lambda_S=0$ in $P(z)$.

The basis states for $Q=0$ and $Y=1$ are,
\begin{equation}
	\phi^+\xi^{+*}, \ \phi^0\xi^0, \ \chi^+\phi^{+*}, \ \chi^0\phi^{0*}, \ S\phi^0.
\end{equation}
Scattering amplitudes involving these states yield four additional distinct eigenvalues of $\mathcal{M}$,
\begin{align}
y_3&=4\lambda_2-\lambda_5, \nonumber \\
y_4&=4\lambda_2+2\lambda_5, \nonumber \\
y_5&=4\lambda_2+4\lambda_5, \nonumber \\
z_a&=4\lambda_a.
\label{eq:eval2}
\end{align}
Scattering amplitudes involving basis states with other values of $Q$ and $Y$ only repeat eigenvalues that have already been found.
Note that by adding the real singlet scalar $S$ we have replaced the two eigenvalues $x_1^{\pm}$ of Refs.~\cite{Aoki:2007ah,Hartling:2014zca} with five new eigenvalues $z_{1,2,3,a,b}$.
We obtain the unitarity bounds by requiring that the absolute value of each of the eigenvalues in Eqs.~(\ref{eq:eval1}) and (\ref{eq:eval2}) be less than $8\pi$. 

The three unitarity constraints $|z_{1,2,3}| < 8 \pi$ can be made more algebraically tractable by replacing them with three equivalent conditions as follows.  First, since Eq.~(\ref{ref:p}) is linear in $\lambda_S$, we can solve the equation $P(z) = 0$ for $\lambda_S$ as a function of the root $z$,
\begin{widetext}
\begin{equation}
\lambda_S(z)= \frac{1}{6}\left(\frac{z}{2} + \frac{2\lambda_a^2(7\lambda_3+11\lambda_4-\frac{1}{8}z)+9\lambda_b^2(3\lambda_1-\frac{1}{8}z)-18\lambda_2\lambda_a\lambda_b}{2(7\lambda_3+11\lambda_4-\frac{1}{8}z)(3\lambda_1-\frac{1}{8}z)-9\lambda_2^2} \right).
\label{eq:lambdaSz}
\end{equation}
\end{widetext}
This function has two poles, across which $\lambda_S(z)$ changes sign.  There are thus three values of $z$ that yield the same value of $\lambda_S$, corresponding to the three roots $z_{1,2,3}$ of the polynomial $P(z)$.  We now require that all three of these roots satisfy $|z_{1,2,3}| < 8 \pi$.  For this to be possible, the two poles in $\lambda_S(z)$ must also lie at $z$ values between $-8\pi$ and $8\pi$.  The positions of these two poles are given by $z = x_1^{\pm}$, where
\begin{eqnarray}
	x_1^\pm &=& 12\lambda_1 + 14\lambda_3 + 22\lambda_4 \nonumber \\
	 && \pm \sqrt{(12\lambda_1-14\lambda_3-22\lambda_4)^2+144\lambda_2^2}.
\end{eqnarray}
Therefore we require $|x_1^{\pm}| < 8 \pi$, reproducing two of the unitarity constraints from the original GM model~\cite{Aoki:2007ah,Hartling:2014zca}.  The third condition restricts $\lambda_S$ to lie in the range for which the three roots of $P(z)$ all lie within $(-8\pi, 8 \pi)$,
\begin{equation}
	\lambda_S^{\rm min} < \lambda_S < \lambda_S^{\rm max},
\end{equation}
where $\lambda_S^{\rm min} = \lambda_S(z = -8\pi)$ and $\lambda_S^{\rm max} = \lambda_S(z = 8 \pi)$ from Eq.~(\ref{eq:lambdaSz}).

To summarize, we will require that the following constraints from perturbative unitarity be satisfied:
\begin{widetext}
\begin{align}
8\pi>&\left| 12\lambda_1+14\lambda_3+22\lambda_4  \pm \sqrt{(12\lambda_1-14\lambda_3-22\lambda_4)^2+144\lambda_2^2} \right| = |x_1^{\pm}|, \nonumber \\
8\pi>&\left|4\lambda_1-2\lambda_3+4\lambda_4 \pm\sqrt{(4\lambda_1+2\lambda_3-4\lambda_4)^2+4\lambda_5^2}\right| = |x_2^{\pm}|, \nonumber \\
8\pi>&|16\lambda_3+8\lambda_4| = |y_1|, \nonumber \\
8\pi>&|4\lambda_3+8\lambda_4| = |y_2|, \nonumber \\
8\pi>&|4\lambda_2-\lambda_5| = |y_3|, \nonumber \\
8\pi>&|4\lambda_2+2\lambda_5| = |y_4|, \nonumber \\
8\pi>&|4\lambda_2+4\lambda_5| = |y_5|, \nonumber \\
8\pi>&|4\lambda_a| = |z_a|, \nonumber \\
8\pi>&|4\lambda_b| = |z_b|, \nonumber \\
\lambda_S<& \frac{1}{6}\left(4\pi + \frac{2\lambda_a^2(7\lambda_3+11\lambda_4-\pi)+9\lambda_b^2(3\lambda_1-\pi)-18\lambda_2\lambda_a\lambda_b}{2(7\lambda_3+11\lambda_4-\pi)(3\lambda_1-\pi)-9\lambda_2^2} \right), \nonumber \\
\lambda_S>& \frac{1}{6}\left(-4\pi+\frac{2\lambda_a^2(7\lambda_3+11\lambda_4+\pi)+9\lambda_b^2(3\lambda_1+\pi)-18\lambda_2\lambda_a\lambda_b}{2(7\lambda_3+11\lambda_4+\pi)(3\lambda_1+\pi)-9\lambda_2^2} \right).
\label{eq:uniconstr}
\end{align}
\end{widetext}

\subsection{Requirement that the scalar potential be bounded from below}

We next examine the constraints on the scalar couplings imposed by requiring that the scalar potential be bounded from below.
The constraints that must be satisfied at tree level for the scalar potential to be bounded 
from below can be determined by considering only the terms that are quartic in the fields, 
because these terms dominate at large field values. Following the approach of Ref.~\cite{Arhrib:2011}, 
we parametrize the potential using the following definitions,
\begin{align}
r&=\sqrt{\mbox{Tr}(\Phi^\dagger \Phi)+\mbox{Tr}(X^\dagger X) +S^2}, \nonumber \\
r^2 \cos^2\gamma \sin^2 \beta &=\mbox{Tr}(\Phi^\dagger \Phi), \nonumber \\
r^2 \sin^2\gamma \sin^2 \beta &=\mbox{Tr}(X^\dagger X), \nonumber \\
r^2 \cos^2 \beta &=S^2, \nonumber \\
\zeta&=\frac{\mbox{Tr}(X^\dagger X X^\dagger X)}{(\mbox{Tr}(X^\dagger X))^2}, \nonumber \\
\omega&=\frac{\mbox{Tr}(\Phi^\dagger \tau^a \Phi \tau^b)\mbox{Tr}(X^\dagger t^a X t^b)}{\mbox{Tr}(\Phi^\dagger \Phi)\mbox{Tr}(X^\dagger X)}.
\label{eq:zetaomega}
\end{align}
Making these substitutions, we can write the quartic part of the potential as
\begin{equation}
V_4=\frac{r^4}{(1+\tan^2\gamma)^2(1+\tan^2 \beta)^2} \mathbf{x}^T A \mathbf{y},
\label{eq:V4}
\end{equation}
where
\begin{equation}
\mathbf{x}=
\begin{pmatrix}
1\\
\tan^2\beta \\
\tan^4\beta
\end{pmatrix} , \qquad
\mathbf{y}=
\begin{pmatrix}
1\\
\tan^2\gamma \\
\tan^4\gamma
\end{pmatrix} , 
\end{equation}
and
\begin{equation} 
A=\begin{pmatrix}
\lambda_S&  2\lambda_S& \lambda_S \\
\lambda_a & \lambda_a+\lambda_b & \lambda_b\\
\lambda_1 &  \lambda_2-\lambda_5\omega& \lambda_3\zeta+\lambda_4
\end{pmatrix}.
\label{eq:bfbA}
\end{equation}
The first fraction in Eq.~(\ref{eq:V4}) is always positive, and grows with the overall field excursion $r$.
The $\mathbf{x}^T A \mathbf{y}$ term in Eq.~(\ref{eq:V4}) can be positive or negative; we require it to be positive to ensure that the potential is bounded from below.  This term can be expressed as a bi-quadratic in $\tan\gamma$ with coefficients being other bi-quadradics in $\tan\beta$. 
A bi-quadratic of the form $a+bz^2+c z^4$ will be positive for all values of $z$ if the following conditions are satisfied:
\begin{equation}
a>0, \qquad c>0, \quad \mbox{and}  \quad b+2\sqrt{ac}>0 .
\end{equation}
In our case this leads to the following constraints on the elements of the matrix $A$ in Eq.~(\ref{eq:bfbA}):
\begin{align}
0&<A_{11}=\lambda_S, \nonumber \\
0&<A_{33}=\zeta \lambda_3+\lambda_4, \nonumber \\
0&<A_{13}=\lambda_S, \nonumber \\
0&<A_{31}=\lambda_1, \nonumber \\
0&<A_{12}+2\sqrt{A_{11}A_{13}}=4\lambda_S, \nonumber \\
0&<A_{32}+2\sqrt{A_{31}A_{33}}=\lambda_2-\omega\lambda_5+2\sqrt{\lambda_1(\zeta\lambda_3+\lambda_4)}, \nonumber \\
0&<A_{21}+2\sqrt{A_{11}A_{31}}=\lambda_a+2\sqrt{\lambda_1 \lambda_S}, \nonumber \\
0&<A_{23}+2\sqrt{A_{13}A_{33}}=\lambda_b+2\sqrt{(\zeta\lambda_3+\lambda_4)\lambda_S}, \nonumber \\
0&<x_i A_{i2}+2\sqrt{x_j A_{j1} x_k A_{k3}} \nonumber \\
& \qquad \qquad \quad =\mathbf{x}^T A \mathbf{e}_2+2\sqrt{(\mathbf{x}^T A \mathbf{e}_1) (\mathbf{x}^T A \mathbf{e}_3)}, \nonumber \\
0&<A_{2i} y_i+2\sqrt{A_{1j} y_j A_{3k} y_k} \nonumber \\
& \qquad \qquad \quad =\mathbf{e}_2^T A \mathbf{y}+2\sqrt{(\mathbf{e}_1^T A \mathbf{y}) (\mathbf{e}_3^T A \mathbf{y})},
\end{align}
where it should be understood that repeated indices are summed over, and 
$\mathbf{e}_i$ is a unit vector with a $1$ in the $i$th component and zeros everywhere else.
 The last two conditions do not provide any new information as they are always satisfied when the others are, but we list them for completeness. 

The ranges of the parameters $\zeta$ and $\omega$ are given, as in the original GM model 
\cite{Hartling:2014zca}, by
\begin{equation}
	\zeta\in \left[ \frac{1}{3},1 \right], \qquad \omega\in \left[ -\frac{1}{4},\frac{1}{2} \right].
\end{equation}
For a given value of $\zeta$, we can write $\omega\in [\omega_-,\omega_+]$, where~\cite{Hartling:2014zca}
\begin{equation}
\omega_\pm(\zeta)=\frac{1}{6}(1-B)\pm\frac{\sqrt{2}}{3}\left[(1-B)\left(\frac{1}{2}+B\right)\right]^{\frac{1}{2}},
\end{equation}
with
\begin{equation}
B\equiv\sqrt{\frac{3}{2}\left(\zeta-\frac{1}{3}\right)}\in [0,1].
\end{equation}
Therefore, we can write  our constraints as follows:
\begin{align}
\lambda_1&>0, \nonumber \\
\lambda_4&>\left\lbrace \begin{matrix}
&-\frac{1}{3}\lambda_3 &\mbox{ for } \lambda_3 \geq 0,\\
&-\lambda_3 &\mbox{ for } \lambda_3 < 0,\\
\end{matrix} \right. \nonumber \\
\lambda_2&>\left\lbrace \begin{matrix}
&\frac{1}{2}\lambda_5-2\sqrt{\lambda_1\left(\frac{1}{3}\lambda_3+\lambda_4\right)} &\mbox{for } \lambda_5 \geq 0, \; \lambda_3\geq 0,\\
&\omega_+(\zeta)\lambda_5-2\sqrt{\lambda_1\left(\zeta\lambda_3+\lambda_4\right)} &  \mbox{for } \lambda_5 \geq 0, \; \lambda_3< 0,\\
&\omega_-(\zeta)\lambda_5-2\sqrt{\lambda_1\left(\zeta\lambda_3+\lambda_4\right)} &\mbox{ for } \lambda_5 <0,
\end{matrix}\right. \nonumber \\
\lambda_a&>-2\sqrt{\lambda_1 \lambda_S}, \nonumber \\
\lambda_b&>\left\lbrace \begin{matrix}
	&-2\sqrt{\left(\frac{1}{3} \lambda_3+\lambda_4\right)\lambda_S} & \mbox{ for } \lambda_3 \geq 0, \\
	&-2\sqrt{\left(\lambda_3+\lambda_4\right)\lambda_S} & \mbox{ for } \lambda_3 < 0, 
	\end{matrix} \right. \nonumber \\
\lambda_S&>0.
\label{eq:bfbconstr}
\end{align}
The first three of these constraints are identical to those in the original GM model, while the last three are new.

We note that the full parameter space of the quartic scalar couplings as allowed by 
perturbative unitarity and the requirement that the scalar potential be bounded from 
below can be covered by scanning over the following ranges.  For the couplings 
$\lambda_1$--$\lambda_5$, the ranges are the same as in the original GM model~\cite{Hartling:2014zca},
\begin{eqnarray}
&&	\lambda_1 \in \left( 0, \frac{\pi}{3} \right), \quad
	\lambda_2 \in \left( -\frac{2\pi}{3}, \frac{2\pi}{3} \right), \quad
	\lambda_3 \in \left( -\frac{\pi}{2}, \frac{3 \pi}{5} \right), \nonumber \\
&&	\lambda_4 \in \left( -\frac{\pi}{5}, \frac{\pi}{2} \right), \quad
	\lambda_5 \in \left( -\frac{8 \pi}{3}, \frac{8 \pi}{3} \right).
\end{eqnarray}
For the new couplings $\lambda_a$, $\lambda_b$, and $\lambda_S$ in the singlet scalar 
dark matter extension of the GM model, the ranges are,\footnote{The upper limits of these 
ranges come from the unitarity constraints in Eq.~(\ref{eq:uniconstr}).  The upper 
limit on $\lambda_a$ comes directly from $|z_a| < 4 \pi$.  The upper limit on 
$\lambda_b$ comes from the upper and lower bounds on $\lambda_S$: for large enough 
$\lambda_b$ these two bounds meet each other, and the least stringent bound on 
$\lambda_b$ comes from taking all other quartic couplings equal to zero in these expressions.  
The upper limit on $\lambda_S$ comes directly from the expression in Eq.~(\ref{eq:uniconstr}), 
which is least stringent when all other quartic couplings are set to zero.

The lower limit on $\lambda_a$ comes from an interplay of the bounded-from-below constraint 
$\lambda_a > -2 \sqrt{\lambda_1 \lambda_S}$ in Eq.~(\ref{eq:bfbconstr}) and the upper bound 
on $\lambda_S$ from 
Eq.~(\ref{eq:uniconstr}) when $\lambda_2 = \lambda_3 = \lambda_4 = \lambda_5 = \lambda_b = 0$ and $\lambda_S = 2 \lambda_1$.  
The lower limit on $\lambda_b$ comes from an interplay of the constraint in Eq.~(\ref{eq:bfbconstr}) 
and the bound on $\lambda_3$ and $\lambda_4$ from $|x_1^{\pm}| < 8 \pi$ in Eq.~(\ref{eq:uniconstr}).  
The least stringent limit occurs when $\lambda_1 = \lambda_2 = \lambda_3 = 0$.
The lower limit on $\lambda_S$ comes trivially from Eq.~(\ref{eq:bfbconstr}).
}
\begin{eqnarray}
&&	\lambda_a \in \left( -\frac{2\pi (3\sqrt{2} - 2)}{7}, 2\pi \right), \quad
	\lambda_b \in \left( -\frac{4\pi}{\sqrt{33}}, \frac{4\pi}{3} \right), \nonumber \\ 
&&	\lambda_S \in \left( 0, \frac{2\pi}{3} \right). 
\end{eqnarray}
Within these ranges, the conditions in Eqs.~(\ref{eq:uniconstr}) and (\ref{eq:bfbconstr}) 
must still be applied and any points in violation discarded.

\subsection{Conditions to avoid alternative minima}

Finally we check that the scalar potential does not contain any deeper minima that spontaneously break the custodial symmetry or that give the singlet a vev.

The constraints on the parameters required to ensure that the desired electroweak-breaking and custodial SU(2)-preserving minimum is the true global minimum were studied for the original GM model in Ref.~\cite{Hartling:2014zca}.  These continue to apply in the singlet-extension that we study here and we implement them as follows.  Using $\zeta$ and $\omega$ from Eq.~(\ref{eq:zetaomega}) and introducing the additional parameters
\begin{eqnarray}
	\sigma &=& \frac{\mbox{Tr}(\Phi^\dagger \tau^a \Phi \tau^b)(UXU^\dagger)_{ab}}{\mbox{Tr}(\Phi^\dagger \Phi)[\mbox{Tr}(X^\dagger X)]^{\frac{1}{2}}}, \nonumber \\
	\rho &=& \frac{\mbox{Tr}(X^\dagger t^a X t^b)(UXU^\dagger)_{ab}}{[\mbox{Tr}(X^\dagger X)]^{\frac{3}{2}}}, \nonumber \\
	x^2 &=& \mbox{Tr}(\Phi^\dagger \Phi), \nonumber \\
	y^2 &=& \mbox{Tr}(X^\dagger X), \nonumber \\
	z^2 &=& S^2,
\end{eqnarray}
the scalar potential can be written as
\begin{align}
V=& \frac{\mu_2^2}{2}x^2+\frac{\mu_3^2}{2}y^2+\frac{\mu_S^2}{2}z^2 +(\lambda_2-\lambda_5\omega)x^2y^2 \notag\\
&  +\lambda_a x^2z^2+\lambda_b y^2z^2  +\lambda_1 x^4+(\lambda_3 \zeta+\lambda_4)y^4    \notag\\
&+\lambda_S z^4 -M_1\sigma x^2y -M_2 \rho y^3.
\end{align}
The parameters $\zeta$, $\omega$, $\sigma$ and $\rho$ capture the dependence on which component(s) of $X$ obtain a vev.  The correct custodial SU(2)-preserving vacuum corresponds to $\zeta = 1/3$, $\omega = 1/2$, $\sigma = \sqrt{3}/4$, and $\rho = 2/\sqrt{3}$~\cite{Hartling:2014zca}.  For a given set of Lagrangian parameters, we check that these values yield the lowest value of the potential $V$ by using the convenient parameterization~\cite{Hartling:2014zca}
\begin{eqnarray}
	\zeta &=& \frac{1}{2} \sin^4 \theta + \cos^4 \theta, \nonumber \\
	\omega &=& \frac{1}{4} \sin^2 \theta + \frac{1}{\sqrt{2}} \sin\theta \cos\theta, \nonumber \\
	\sigma &=& \frac{1}{2 \sqrt{2}} \sin\theta + \frac{1}{4} \cos\theta, \nonumber \\
	\rho &=& 3 \sin^2 \theta \cos\theta,
\end{eqnarray}
and scanning over $\theta \in [0, 2\pi)$.

We then check that the potential does not have any deeper minima in which $S$ gets a vev.  If $\mu_S^2$, $\lambda_a$ and $\lambda_b$ are all positive, then $S$ cannot get a vev.  We only have to worry about this possibility if one or two of these parameters are negative (all three cannot be negative because we require $m_S^2 = \mu_S^2 + 2 \lambda_a v_{\phi}^2 + 6 \lambda_b v_{\chi}^2 > 0$).  Taking $\partial V/\partial z = 0$ yields two possible extrema,
\begin{eqnarray}
	z &=& 0, \nonumber \\
	z^2 &=& - \frac{1}{4 \lambda_S} ( \mu_S^2 + 2 \lambda_a x^2 + 2 \lambda_b y^2).
	\label{eq:zsolutions}
\end{eqnarray}
We then take $\partial V/\partial x = 0$ and $\partial V/\partial y = 0$, plug in each of the two solutions for $z$ from Eq.~(\ref{eq:zsolutions}), solve for the possible values of $x$ and $y$ in each case, and then plug these back into $V$ to obtain the depth of the potential at each extremum.  Points are discarded if a minimum with $z \neq 0$ is deeper than the desired one with $z = 0$.

\section{THERMAL RELIC DENSITY}
\label{sec:freezeout}

We now turn to constraints from the dark matter relic abundance.  We assume that the scalar dark matter candidate $S$ 
constitutes all of the dark matter.  We will use the observed relic density to fix a combination of $\lambda_a$ and $\lambda_b$.  
We show that the direct detection constraints restrict the dark matter mass to be near half the Higgs mass around $62$ GeV
or above approximately $120$ GeV.

\subsection{Thermal Freezeout}

The relic density of $S$ through thermal freeze-out in the early universe is determined 
by the annihilation cross section for $SS \to {\rm anything}$.   We calculate 
the thermally averaged cross section as a function of temperature using \cite{Gondolo:1990dk,Edsjo:1997bg}:
\begin{align}
&\langle \sigma_{12\rightarrow34}v_{\rm rel} \rangle 
 = \frac{g_1 g_2 T}{32\pi^4 n_1^{\rm eq}n_2^{\rm eq}} \\
& \times \int_{4m_S^2}^{\infty} \sigma_{12\rightarrow34}\left[s-4m_S^2 \right]^2\sqrt{s} K_1\left(\frac{\sqrt{s}}{T}\right)ds, \nonumber
\end{align}
where $s=(p_1+p_2)^2$ is the usual Mandelstam variable, $g_1=g_2 \equiv g_S = 1$ is the number of 
internal degrees of freedom of $S$, $v_{\rm rel}$ is the relative velocity of particles $1$ and $2$, and $K_1$ is the modified Bessel function of the second kind of order 1.
We use the thermally averaged total annihilation cross section as input for the usual Boltzmann equation 
\cite{Kolb:1990vq,Gondolo:1990dk,Edsjo:1997bg}:
\begin{align}
\frac{dn_S}{dt}+3Hn_S=-\langle \sigma v_{\rm rel} \rangle \left[n_S^2 - \left(n_S^{\rm eq}\right)^2 \right],
\end{align}
where $H$ is the Hubble parameter and $n_S$ is the number density of $S$. Here $n_S^{\rm eq}$ 
is the equilibrium number density of $S$ and is given by \cite{Gondolo:1990dk,Edsjo:1997bg}:
\begin{align}
n_S^{\rm eq} &= \frac{g_S}{(2\pi^3)}\int e^{-E_S/T}d^3p_i\notag\\
&=\frac{g_S m_S^2 T}{2\pi^2}K_2(m_S/T)\\
&\approx g_S\left(\frac{m_S T}{2\pi}\right)^{3/2}e^{-m_S/T},
\end{align}
where $K_2$ is the modified Bessel function of the second kind of order 2 and where the 
approximation holds for when $m_S\gg T$. We then solve this equation numerically to obtain the
number density today which translates to a value for the relic abundance.

We need to include all final states arising from $SS$ annihilation into SM
particles and various other scalar final states appearing in the model. 
We group them by final state.

\subsubsection{$f \bar f$, $VV$, and $H_3V$ final states}

\begin{figure}
   \includegraphics[width=0.4\textwidth,clip]{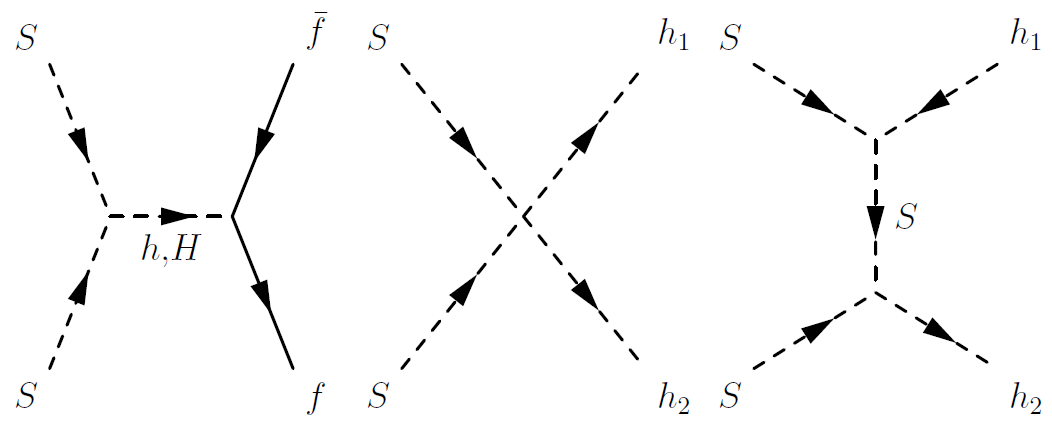}
   \caption{Feynman diagrams for $SS$ annihilation. In the first, the final state can be any allowed pair of SM particles or GM scalars. However, in the second and third diagrams, only scalar final states can appear. Crossed diagrams (not shown) are also included in the calculation.}
   \label{fig:cs}
\end{figure}

We begin with the final states for which the $SS$ annihilation proceeds by the $s$-channel exchange of $h$ and $H$ bosons only (the first diagram in Fig.~\ref{fig:cs}).  We  write
this annihilation cross sections by incorporating the expression for
the SM Higgs decay width, setting the Higgs mass to the center of mass energy.
For decays to $AB = f \bar f$ or $VV$, the resulting expression is: 
\begin{eqnarray}
	\sigma  v_{\rm rel} & = & \frac{2}{\sqrt{s}} \left(\frac{g_{SSh}g_{hAB}}{s-m_{h}^2}+\frac{g_{SSH}g_{HAB}}{s-m_{H}^2}\right)^2 \frac{1}{g^2_{h_{SM}AB}} \nonumber \\
& & \qquad \times	\Gamma(m_{h_{SM}}=\sqrt{s},h_{SM}\rightarrow AB),
\end{eqnarray}
where $\Gamma$ is the decay width of a SM
 Higgs boson with a mass of $\sqrt{s}$ into final state $AB$. 
 This decay width is calculated using the usual SM formulas.

We must also include the final state with one $H_3$ scalar and one vector boson.
The cross section of this process when $\sqrt{s}>m_3+m_V$ is:  
\begin{eqnarray}
\sigma v_{\rm rel} & = & \frac{(s^2-2s(m_3^2+m_V^2)+(m_3^2-m_V^2)^2)^{\frac{3}{2}}}{8\pi s^2 m_W^2} \nonumber\\
& & \qquad \times  \left(\frac{g_{SSh}g_{hH_3V}}{s-m_{h}^2}+\frac{g_{SSH}g_{HH_3V}}{s-m_{H}^2}\right)^2.
\end{eqnarray}
When $\sqrt{s}<m_3+m_V$, we include the offshell process $SS\rightarrow V^*H_3$, whose cross section is given by
\begin{equation}
\sigma v_{\rm rel}=\left(\frac{g_{SSh}g_{hH_3V}}{s-m_{h}^2}+\frac{g_{SSH}g_{HH_3V}}{s-m_{H}^2}\right)^2\delta_V\frac{3 m_V^2 }{8 \pi^3 v^2}G_{ij},
\end{equation}
where~\cite{GMCALC,Djouadi:1995gv,Akeroyd:1998dt}
\begin{widetext}
\begin{align}
G_{ij}=&\frac{1}{4}\left[ { 2(-1+k_j-k_i)\sqrt{\lambda_{ij}}\left(\frac{\pi}{2}+\arctan\left(\frac{k_j(1-k_j+k_i)-\lambda_{ij}}{(1-k_i)\sqrt{\lambda_{ij}}}\right)\right) }\right. \notag\\
&\qquad \left. { +(\lambda_{ij}-2k_i)\log k_i+\frac{1}{3}(1-k_i)\left(5(1+k_i)-4k_j+\frac{2\lambda_{ij}}{k_j}\right) }\right], 
\end{align}
\end{widetext}
with
\begin{equation}
\delta_W=\frac{3}{2}, \qquad
\delta_Z=3\left(\frac{7}{12}-\frac{10}{9}s_W^2+\frac{40}{27}s_W^4 \right).
\end{equation}
Here  $k_i= m_3^2 /s$, $k_j=m_{V}^2 /s$, $s_W$ is the sine of the weak mixing angle, and $\lambda_{ij}=-1+2k_i+2k_j-(k_i-k_i)^2 $.

\subsubsection{Final states involving $H_3$ or $H_5$ pairs}

We now compute the $SS$ annihilation cross sections into final states that consist of neutral or charged $H_3$ and $H_5$ scalars. 
The possible final states considered here are:
\begin{equation}
(H_3^0,H_3^0),(H_3^+,H_3^-),(H_5^0,H_5^0),(H_5^+,H_5^-),(H_5^{++},H_5^{--}).
\label{eqn:lmpairs}
\end{equation}
Since the two particles in each of these final states have the same mass, we will label it as $m_i$. Annihilation into these final states proceeds via $s$-channel and four-point diagrams (the first two diagrams in Fig.~\ref{fig:cs}).
The cross section for final-state particles $h_1$ and $h_2$ is given by:
\begin{align}
& \sigma v_{\rm rel}  = \delta_{12} \frac{(s-4m_{h_1}^2)^{\frac{1}{2}}}{16 \pi s^{\frac{3}{2}}} \\
& \times \left(g_{SSh_1h_2}+\frac{g_{SSh}g_{hh_1h_2}}{s-m_{h}^2}+\frac{g_{SSH}g_{Hh_1h_2}}{s-m_{H}^2}\right)^2, \nonumber
\end{align}
where $\delta_{12} = 1/2$ for identical final-state particles $h_1 = h_2$ and $\delta_{12} = 1$ for non-identical final-state particles.

\subsubsection{$(h,H)$, $(h,h)$ and $(H,H)$ final states}

The cross sections with $(h,H)$, $(h,h)$ or $(H,H)$ in the final state 
proceed via $s$-channel, four-point, and $t$- and $u$-channel diagrams (all the diagrams in Fig.~\ref{fig:cs} plus the crossed diagram). For non-identical $h_i,h_j$ in the final state, we obtain:
\begin{align}
\sigma v_{\rm rel} =\frac{\sqrt{s^2+(m_{h_j}^2-m_{h_i}^2)^2-2s(m_{h_j}^2+m_{h_i}^2)}}{16\pi s^2}g(s),
\end{align}
and for identical particles in the final state, we obtain:
\begin{align}
\sigma v_{\rm rel} =\frac{\sqrt{s-4m_{h_i}^2}}{32\pi s^{3/2}}g(s),
\end{align}
where $g$, $a$, $b$, $c$, and $d$ are defined as follows:
\begin{align}
g(s) =& 2a^2+\frac{b^2}{c^2(c^2-d^2)}+\frac{b(b-4ac^2)}{c^3d}\tanh^{-1}\left(\frac{d}{c}\right), \\
a=&g_{SSh_ih_j} +
\sum_k \frac{g_{SSh_k}g_{h_kh_ih_j}}{s-m_{h_k}^2},\\
b=&4g_{SSh_i}g_{SSh_j}(s-(m_{h_i}^2+m_{h_j}^2)),\\
c^2=&(s-m_{h_j}^2-m_{h_i}^2)^2,\\
d^2=&\left(1 - \frac{4m_S^2}{s}\right)\left((s-m_{h_j}^2+m_{h_i}^2)^2 - 4s m_{h_i}^2\right),
\end{align}
where the sum in $a$ runs over $h_k = h, H$.

\subsection{Imposing Relic Density}

Here we give details on how we impose the relic density as a constraint on $\lambda_a$ 
and $\lambda_b$. We first note that the thermally averaged cross section is a strictly increasing function of  
$|\lambda_a|$ and $|\lambda_b|$. In our numerical scans, 
we would like to be able to randomly select a particular linear combination of the couplings $\lambda_a$ and $\lambda_b$ and then scale them both until the correct relic density is obtained. After generating a scan point in the original GM model (see next section for details), we select the values of $\lambda_a$, $\lambda_b$ and $\mu_S^2$ as follows:
\begin{itemize}
  \item First, generate a random angle $\theta_\lambda \in \left[ -\frac{\pi}{2},\frac{\pi}{2} \right]$;
  \item Randomly select either the positive or negative solution for $\lambda_a$;
  \item Set $\lambda_b=\lambda_a \tan \theta_\lambda$;
  \item Generate a random mass $m_S > 0 $ GeV;
  \item Find a value of $\lambda_a$ that yields a relic density of $0.1064\leq \Omega_{DM} h^2 \leq 0.1176$~\cite{Komatsu:2010fb};
  \item Once the value of $\lambda_a$ is found we can find $\lambda_b$ and, in turn, $\mu_S^2$ using Eq.~(\ref{eq:m_s}).
\end{itemize}

The first three steps allows us to select a particular linear combination of 
$\lambda_a$ and $\lambda_b$. Generating $m_S$ directly lets us avoid unphysical negative $m_S^2$ values regardless of the actual values of $\lambda_a$ and $\lambda_b$. Finally, numerically searching for the correct value of $\lambda_a$ is straightforward because $\sigma v_{\rm rel}$ is an increasing function of $\lambda_a$ and $\sigma v_{\rm rel}\rightarrow \infty$ when $\lambda_a\rightarrow \infty$ and $\sigma v_{\rm rel}\rightarrow 0$ when $\lambda_a\rightarrow 0$.

\section{Numerical Scan Procedure}
\label{sec:Scan}

To map out the allowed parameter space, we  perform numerical scans. 
In these scans, we start by imposing the theoretical constraints and the relic density constraint. 
We then check whether the points pass the remaining experimental constraints. 
In the plots that follow, points that pass the experimental constraints will be blue, while points 
that fail at least one constraint will be red. 

For the dimensionful parameters, the ranges we scan over are:
\begin{itemize}
	\item $-4\times 10^4$~GeV$^2 \leq \mu_3^2 \leq 10^6$~GeV$^2$;
	\item $0\leq M_1 \leq \max(3500~{\rm GeV}, 3.5\sqrt{|\mu_3^2|})$;
	\item $|M_2| \leq \max(250~{\rm GeV},1.3 \sqrt{|\mu_3^2|})$, with either sign allowed;
	\item $0\leq m_S \leq 1000$~GeV, $0\leq m_S \leq 125$~GeV, or $56\leq m_S \leq 63$~GeV (see text 
	below for explanation of these three regions).
\end{itemize}
The ranges for $M_1$ and $M_2$ are chosen to minimize the number of points generated which 
fail the theory constraints while still scanning the whole parameter space. The mass parameters $\mu_3^2$ and $m_S$ do not have upper bounds so we impose arbitrary bounds for the purpose of the scan.  We perform a scan with $0 \leq m_S \leq 1000$~GeV 
in order to obtain a general picture of the parameter space, one with $0 \leq m_S \leq 125$~GeV  
in order to obtain higher statistics in the interesting low-$m_S$ region, and finally a smaller 
dedicated scan with $56 \leq m_S \leq 63$~GeV to further investigate the Higgs pole region.

From these values, we calculate $\lambda_1$, $\mu_2^2$ and $\mu_S^2$, and all the masses and couplings, and then use the relic density to fix a random linear combination of $\lambda_a$ and $\lambda_b$.

\section{Direct and indirect collider constraints on the GM model}
\label{sec:collider}

Very low masses for $H_3^{0, \pm}$ and $H_5^{0, \pm, \pm\pm}$ can have a substantial effect on the dark matter relic abundance through annihilations into pairs of these scalars.  We constrain these masses using direct experimental search limits as follows.
LHC limits on anomalous like-sign dimuon production~\cite{ATLAS:2014kca} set a lower bound on the mass of a doubly-charged scalar decaying to like-sign $W$ boson pairs.  This was studied in Ref.~\cite{Kanemura:2014ipa} for the Higgs Triplet Model~\cite{HTM} and recast into the GM model in Ref.~\cite{Logan:2015xpa}.  This yields a lower bound $m_5 \geq 76$~GeV, so long as $H_3$ is heavier than $H_5$ so that decays $H_5^{\pm\pm} \to W^{\pm} H_3^{\pm}$ do not compete with the decays into like-sign $W$ pairs.
Searches for a charged Higgs boson at the CERN Large Electron-Positron (LEP) collider~\cite{Searches:2001ac} exclude charged Higgs masses below 78~GeV, assuming that the charged Higgs decays entirely into a combination of $\tau\nu$ and $cs$ final states.  This limit can be applied to $H_3^{\pm}$ so long as decays $H_3^{\pm} \to W^{\pm} H_5^0, Z H_5^{\pm}$ do not compete with the decays to fermions.  This holds when $H_5$ is heavier than $H_3$.  We therefore impose the lower bounds
\begin{equation}
	m_3 \geq 76~{\rm GeV}, \qquad \qquad m_5 \geq 76~{\rm GeV}.
\end{equation}

Low $H_3^{\pm}$ masses can also be constrained from their effect on the loop-induced decay of $b \to s \gamma$.  
We use the ``loose'' constraint determined for the GM model in Ref.~\cite{Hartling:2014aga}, which is based on 
an experimental average from the Heavy Flavour Averaging Group~\cite{Beringer:1900zz,Asner:2010qj} and a 
theoretical prediction from the public code SuperIso~v3.3~\cite{Mahmoudi:2007vz}.  The $b \to s \gamma$ 
constraint sets a maximum value of $v_{\chi}$ as a function of $m_3$. Although this could potentially 
be constraining, all points in our numerical scan satisfied this constraint.

\section{Constraints from dark matter}
\label{sec:darkmatter}

\subsection{Dark Matter Direct Detection}
\label{sec:DM_direct_contraint}

\begin{figure}[t]
   \centering
   \includegraphics[width=0.3\textwidth]{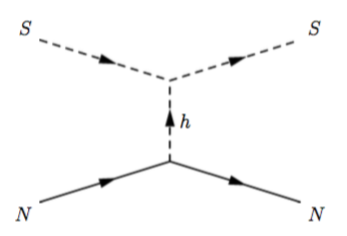}
   \caption{Feynman diagram for direct detection via t-channel Higgs exchange in $SN\to SN$ scattering.  There is a second diagram in which $h$ is replaced by $H$.}
   \label{fig:directde}
\end{figure}

When a dark matter particle is in close proximity with a nucleon, there may be a scattering 
via the t-exchange of a Higgs boson. This transfer of momentum can be detected from the nucleon recoil
so that experimental 
limits can be used to constrain our model. In our model, this process proceeds via exchange of a 
virtual $h$ or $H$ as shown in Fig. \ref{fig:directde}.

The spin-independent cross section for the scattering of a scalar dark matter particle $S$ off of a single nucleon is given by
\begin{equation}
	\sigma=\left(\frac{g_{SSh}c_{\alpha}}{m_{h}^{2}c_{H}} + \frac{g_{SSH}s_{\alpha}}{m_{H}^{2}c_{H}}\right)^{2} \frac{f_N^2 m_{N}^{4}}{4\pi(m_{N}+m_{S})^{2}v^{2}},
\label{eq:sigmaSI}
\end{equation}
where we neglect the momentum transfer relative to the $h$ or $H$ mass, $c_H$ was defined in Eq. (\ref{eq:cHsH}), $c_\alpha \equiv \cos \alpha$, $s_\alpha \equiv \sin \alpha$, 
and $f_{N}$ is the nucleon vertex factor \cite{Cline:2013gha}, 
\begin{equation}
f_{N}=\sum_{q}f_{q}=\sum_{q}\frac{m_{q}}{m_{N}}<N|\bar{q}q|N>,
\label{eq:fN}
\end{equation}
where the sum is over all quark flavours, and the Feynman rule for the Higgs-nucleon vertex is $-i f_N m_N/v$.
We follow Ref.~\cite{Cline:2013gha} in using $f_N=0.30\pm 0.03$ and $m_N=(m_n+m_p)/2=938.95$ MeV.

\begin{figure}[t]
\resizebox{0.45\textwidth}{!}{\includegraphics{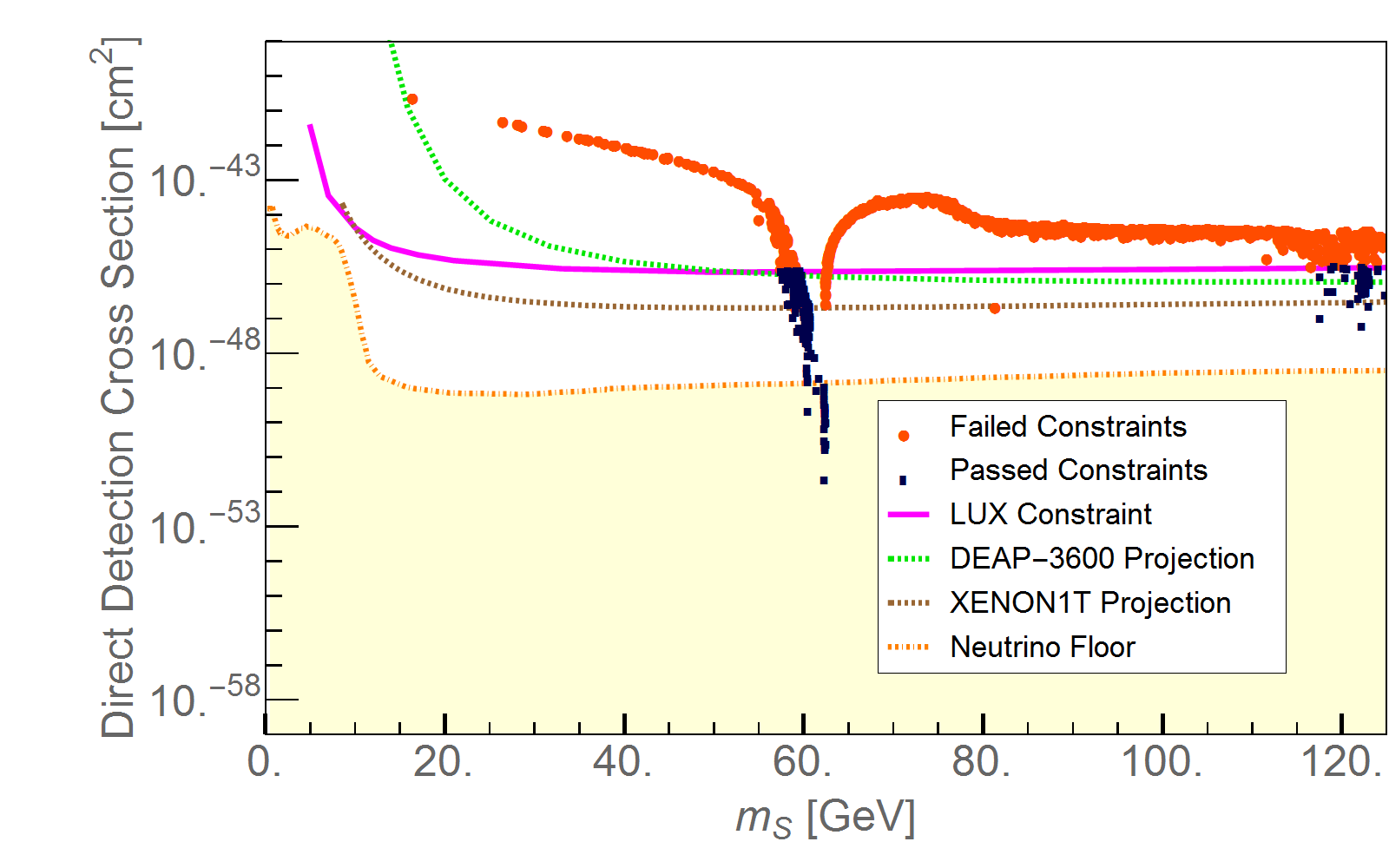}}
\resizebox{0.482\textwidth}{!}{\includegraphics{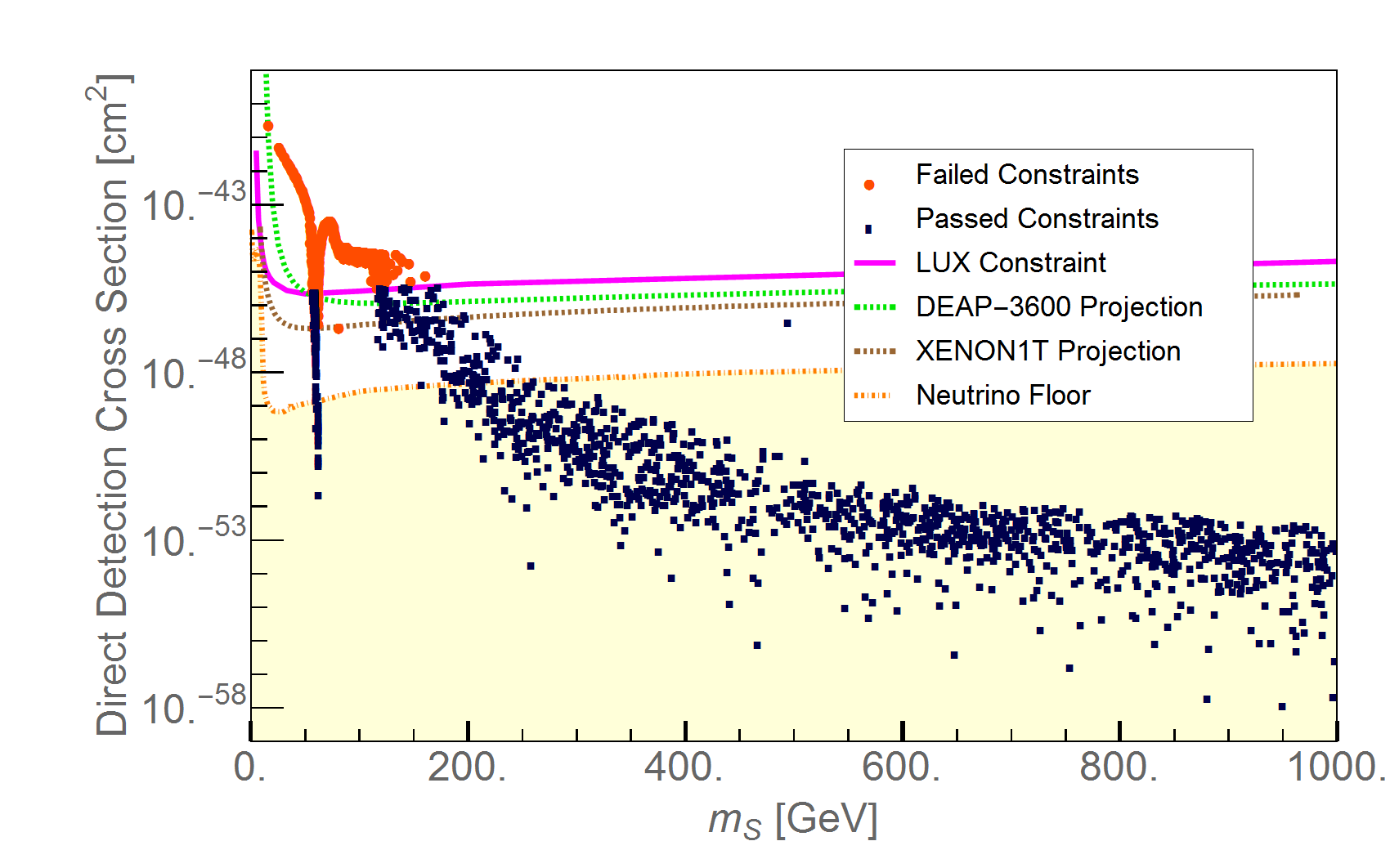}}
\caption{Direct detection cross section as a function of $m_S$.  
The top plot is a zoom of the low-$m_S$ region while the bottom plot shows the 
full $m_S$ range scanned.  The red (grey) points are eliminated by direct or indirect dark matter
detection measurements and the blue (black) points are the remaining ones that are allowed.
On the right side of the figures, starting at the top are 
the exclusion limits from LUX (solid magenta) \cite{Akerib:2016vxi}, 
as well as projected sensitivities of DEAP-3600 (dashed green)~\cite{Amaudruz:2014nsa}, 
and XENON1T (dashed brown)~\cite{Cushman:2013zza}. The yellow shaded region below the lowest line
is the coherent neutrino scattering background (``neutrino floor")~\cite{Billard:2013qya}.
Data files were taken from the DMTools website~\cite{DMtools}.}
\label{fig:dirdet}
\end{figure}

In Fig.~\ref{fig:dirdet} we illustrate the effect of the direct detection constraints on our model.  The scan points shown are those that satisfy the theoretical constraints and yield the correct dark matter relic abundance.  The blue points satisfy the constraints from the dark matter direct detection experiments as well as limits from indirect detection (see next subsection), while the red points fail those constraints.  The current most stringent direct-detection cross section limit comes from the LUX experiment~\cite{Akerib:2016vxi} and is shown as the solid magenta line in Fig.~\ref{fig:dirdet}.  As can be seen, this constraint is responsible for excluding the great majority of the red (excluded) points in our scan, except for a small collection of points on the higher-mass side of the Higgs pole at $m_h/2 = 62.5$~GeV.
We also show the projected limits from DEAP-3600~\cite{Amaudruz:2014nsa} (dotted green)
and XENON1T \cite{Cushman:2013zza} (dotted brown), as well as the ``neutrino floor'' (yellow shaded region) below which coherent neutrino scattering becomes an irreducible background to the dark matter direct detection experiments~\cite{Billard:2013qya}.

\subsection{Dark matter indirect detection}
\label{sec:indirdet}

Dwarf spheroidal satellite galaxies (dSphs) are typically dark matter dominated so are 
a good place to study dark matter. The Fermi collaboration has acquired 6 years worth of 
data observing 15 dSphs and have released bounds for WIMP dark matter annihilation based on 
their gamma ray flux. They considered the following representative final states for the 
dark matter annihilation: $e^+e^-$, $\mu^+\mu^-$, $\tau^+\tau^-$, $u\bar{u}$, $b\bar{b}$, and $W^+W^-$ \cite{FermiLAT}.

\begin{figure}[t]
   \centering
   \includegraphics[width=0.45\textwidth]{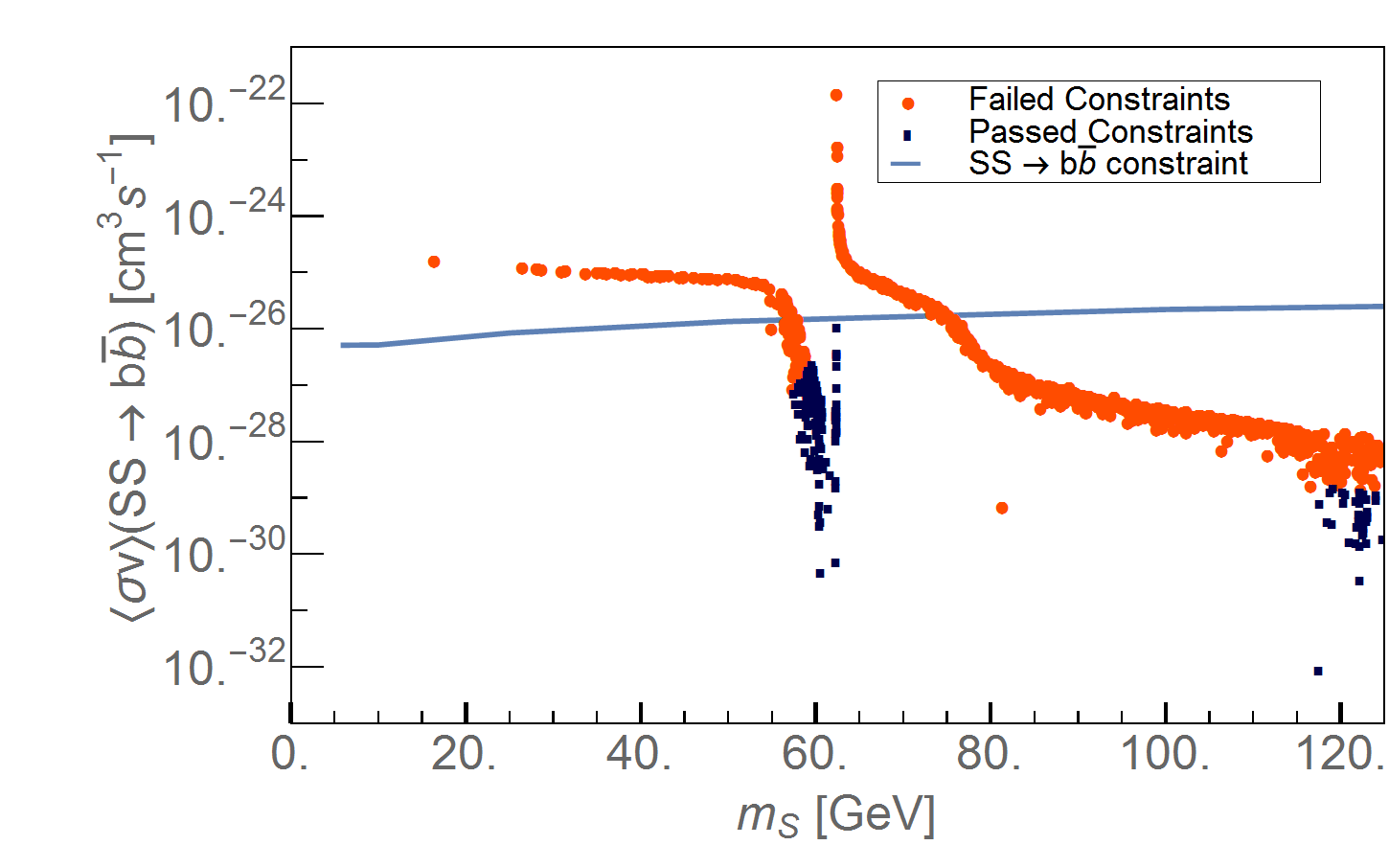}
   \caption{Present-day annihilation cross section for $SS \to b \bar b$ as a function of $m_S$.  The blue (black) points are allowed while the red (grey) points 
   are excluded by direct detection or dwarf spheroidal galaxy constraints.
   Points above the blue line are excluded by the dSphs $b\bar{b}$ constraint from the Fermi satellite~\cite{FermiLAT}.
   Fermi gives the thermally averaged cross section while we use the low velocity approximation and take $s=4m_S^2$.}
   \label{fig:dSph}
\end{figure}

We can translate the Fermi bounds into constraints on our model by considering the branching ratio of 
the singlet annihilation to these final states. Although all of the final states are 
considered in our analysis, the strongest constraint comes from the $b\bar{b}$ final 
state for singlet scalar masses just below half the $h$ mass. 
Figure~\ref{fig:dSph} shows the results of applying the  
$b\bar{b}$ constraint to the scan points. As can be seen, 
there is a sharp dip in the cross section followed by a sharp peak. The dip can be understood 
as coming from having to lower the values of $\lambda_a$ and $\lambda_b$ near the Higgs resonance, 
therefore lowering the $g_{SSh}$ coupling, in order to obtain the correct relic density. As 
the singlet mass approaches the Higgs pole, the thermal distribution during freeze-out pushes the center 
of mass energy above the pole. This results in increased values of $\lambda_a$ and $\lambda_b$ 
to obtain the correct relic density. However, since the temperature of dark matter is much lower 
today (we use the approximation that $v_{\rm rel}=0$), the increased coupling appears at a center 
of mass energy closer to the Higgs resonance and creates this peak.  The indirect detection constraint thereby excludes a small collection of points on the heavier side of the $h$ pole dip in Fig.~\ref{fig:dirdet} that are not yet excluded by direct detection.

\section{Contraints from Higgs boson properties}
\label{sec:consequences}

\subsection{Higgs Invisible Width}
\label{sec:invisible_width}

When $m_{S}<m_{h,H}/2$, the decay of the Higgs boson to two dark matter candidates is kinematically accessible. 
For convenience we define:
 \begin{align}
 \kappa_{f}^h &= \frac{g_{hf\bar{f}}}{g^{SM}_{hf\bar{f}}}=\frac{c_\alpha}{c_H},\nonumber \\
 \kappa_{V}^h &= \frac{g_{hVV}}{g^{SM}_{hVV}}=c_\alpha s_H - \sqrt{\frac{8}{3}}s_\alpha s_H, \nonumber \\
 \kappa_{\gamma}^h &= \left[\frac{\Gamma(h\rightarrow \gamma\gamma)}{\Gamma^{SM}(h\rightarrow \gamma\gamma)}\right]^{1/2}.
 \label{eq:kappas}
 \end{align} 
Note that $\kappa_f^h$ is the same for all fermions and $\kappa_V^h$ is the same for $V=Z$ and $W^\pm$. 
$\Gamma(h\rightarrow \gamma\gamma)$ receives contributions from $H_3^+$, $H_5^+$, and $H_5^{++}$ 
in addition to the modified $ht \bar t$ and $hWW$ couplings. The expression for the width of this process is
\begin{equation}
\Gamma(h \to {\rm inv})=\frac{g_{SSh}^{2}}{32m_{h}^{2}\pi}\sqrt{m_{h}^{2}-4m_{S}^{2}}.
\end{equation}

The most stringent LHC constraint on invisible Higgs decay comes from Higgs production in vector boson fusion (VBF). 
To compare with experiment, we therefore consider the ratio~\cite{Aad:2015pla}
\begin{equation}\label{eqn:GMbr}
\frac{\sigma_{VBF}{\rm BR}_{\rm inv}}{\sigma_{VBF}^{SM}}=\frac{(\kappa_{V}^{h})^{2}\Gamma(h\rightarrow {\rm inv})}{\Gamma_{\rm tot}}<0.29,
\end{equation}
written in terms of the vector boson fusion (VBF) production cross section and the invisible branching ratio. In the total width $\Gamma_{\rm tot}$ of $h$ we include decays to $b\bar b$, $c \bar c$, $\tau\tau$, $gg$, $WW^*$, $ZZ^*$, $\gamma\gamma$, and $SS$ as computed above.

\begin{figure}[t]
   \centering
   \includegraphics[width=0.45\textwidth]{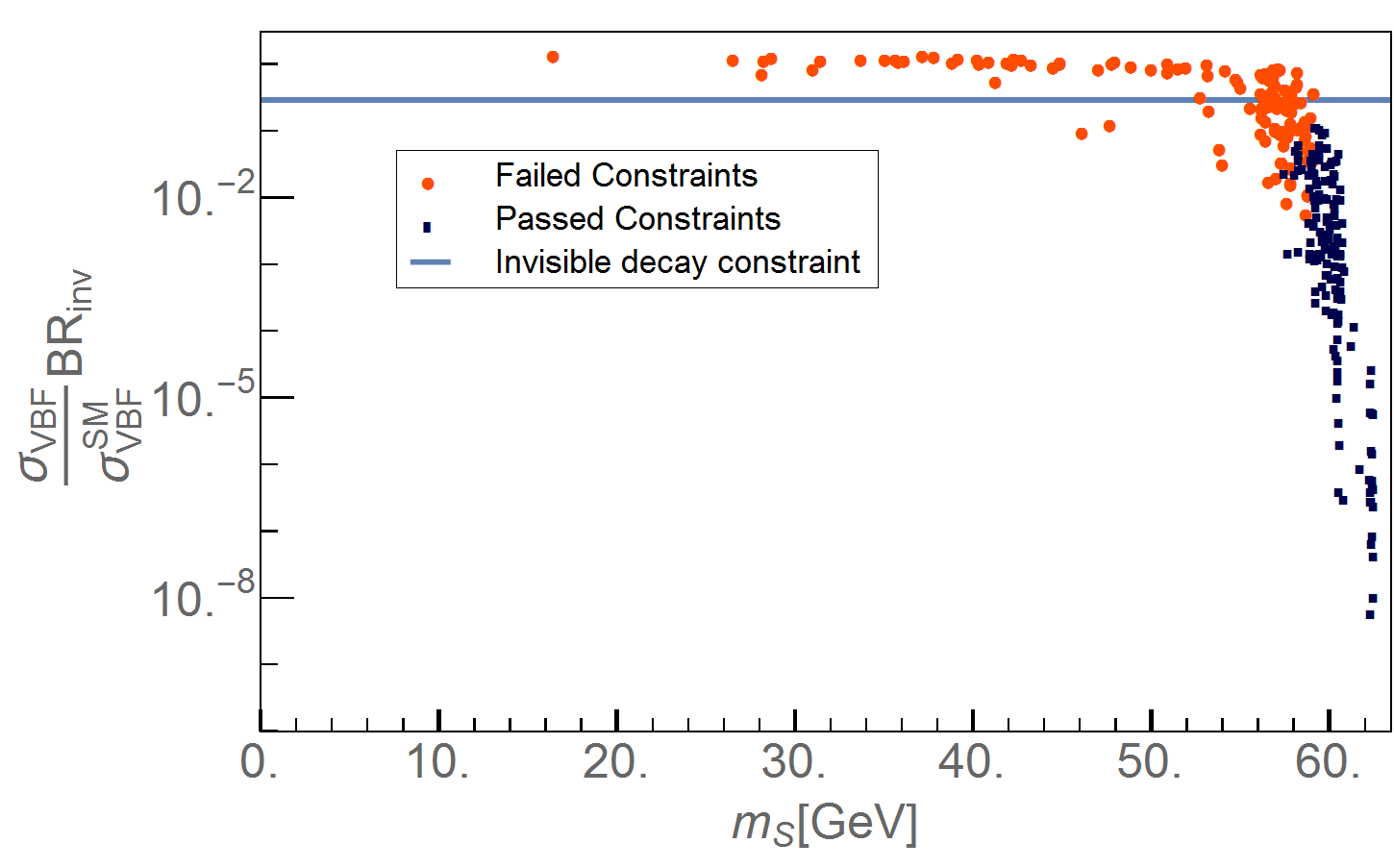}
   \caption{Cross section times branching ratio for $h$ production in vector boson fusion followed by invisible decays to $SS$, normalized to the SM cross section, as a function of $m_S$. The blue (black) points are allowed while the red (grey) points are 
   excluded by direct detection or dwarf spheroidal galaxy constraints. Points above the blue line are excluded by the ATLAS limit on $\sigma_{VBF} {\rm BR}_{\rm inv}/\sigma^{SM}_{VBF}$~\cite{Aad:2015pla} .}
   \label{fig:invGMS}
\end{figure}

The ratio in Eq.~(\ref{eqn:GMbr}) is shown in Fig.~\ref{fig:invGMS}, plotted against $m_{S}$ in  
the kinematically allowed region. The experimental constraint of $\sigma_{VBF} {\rm BR}_{\rm inv}/\sigma^{SM}_{VBF} < 0.29$ \cite{Aad:2015pla} 
is shown as the horizontal blue line.  The bound from invisible Higgs decays is currently not as strong as the constraints from direct detection of dark matter.

\subsection{Higgs Couplings and Signal Strength}
\label{sec:invisible_width}

We finally apply the latest combined measurements of Higgs couplings from CMS and ATLAS from Run 1 
of the LHC~\cite{ATLAS:2016HiggsSignals} to our model.  In this section we discard the points that are 
excluded by dark matter direct detection or indirect detection constraints.  We will find that the 
Higgs coupling measurements exclude a significant fraction of the remaining points, in particular those for 
which the $h$ coupling to fermion or vector boson pairs is sufficiently different from the SM.  

We write the cross sections and branching ratios in terms of the appropriate SM values and the $\kappa$ factors defined in Eq.~(\ref{eq:kappas}) as follows:
\begin{align}
\sigma(AB\rightarrow h) &= (\kappa^h_{AB})^2 \sigma^{SM}(AB\rightarrow h), \nonumber \\
{\rm BR}(h \rightarrow AB) &= (\kappa^h_{AB})^2 \frac{\Gamma^{SM}_{\rm tot}}{\Gamma_{\rm tot}}{\rm BR}^{SM}(h \rightarrow AB),
\end{align}
where $\kappa^h_g = \kappa^h_f$.

We compute a $\chi^2$ using the ATLAS+CMS combined results for the LHC Run 1 Higgs properties 
from Table~9 and the correlation matrix from Fig.~28 of Ref.~\cite{ATLAS:2016HiggsSignals}. The inputs we use in this 
analysis are: $\sigma(gg\rightarrow h\rightarrow ZZ)$, $\sigma_{\rm VBF}/\sigma_{gg\rm F}$, 
$\sigma_{Wh}/\sigma_{gg\rm F}$, $\sigma_{Zh}/\sigma_{gg\rm F}$, $\sigma_{tth}/\sigma_{gg\rm F}$, 
${\rm BR}^{WW}/{\rm BR}^{ZZ}$, ${\rm BR}^{\gamma\gamma}/{\rm BR}^{ZZ}$, ${\rm BR}^{\tau\tau}/{\rm BR}^{ZZ}$, 
and ${\rm BR}^{bb}/{\rm BR}^{ZZ}$. We start by symmetrizing the uncertainties for a given observable 
by taking the root mean square of the asymmetric uncertainties. We then construct the 
variance matrix $V$ from these symmetrized uncertainties and the correlation matrix and define:
\begin{equation}
\chi^2=(\mathbf{x}-\mathbf{y})^TV^{-1}(\mathbf{x}-\mathbf{y}),
\end{equation}
where $\mathbf{x}$ is a vector of the experimental best fit values and $\mathbf{y}$ is a vector of calculated 
values using the kappas and the SM predictions from Table~9 of Ref.~\cite{ATLAS:2016HiggsSignals}. 
Using the SM predictions for $\mathbf{y}$ we get:
\begin{equation}
\chi^2_{SM}=30.0448.
\end{equation}
We will consider a point in our model to be consistent with the experimental measurements of these observables if:
\begin{align*}
\Delta \chi^2 \equiv |\chi^2-\chi^2_{SM}|<4.
\end{align*}

\begin{figure}[t]
   \centering
   \includegraphics[width=0.45\textwidth]{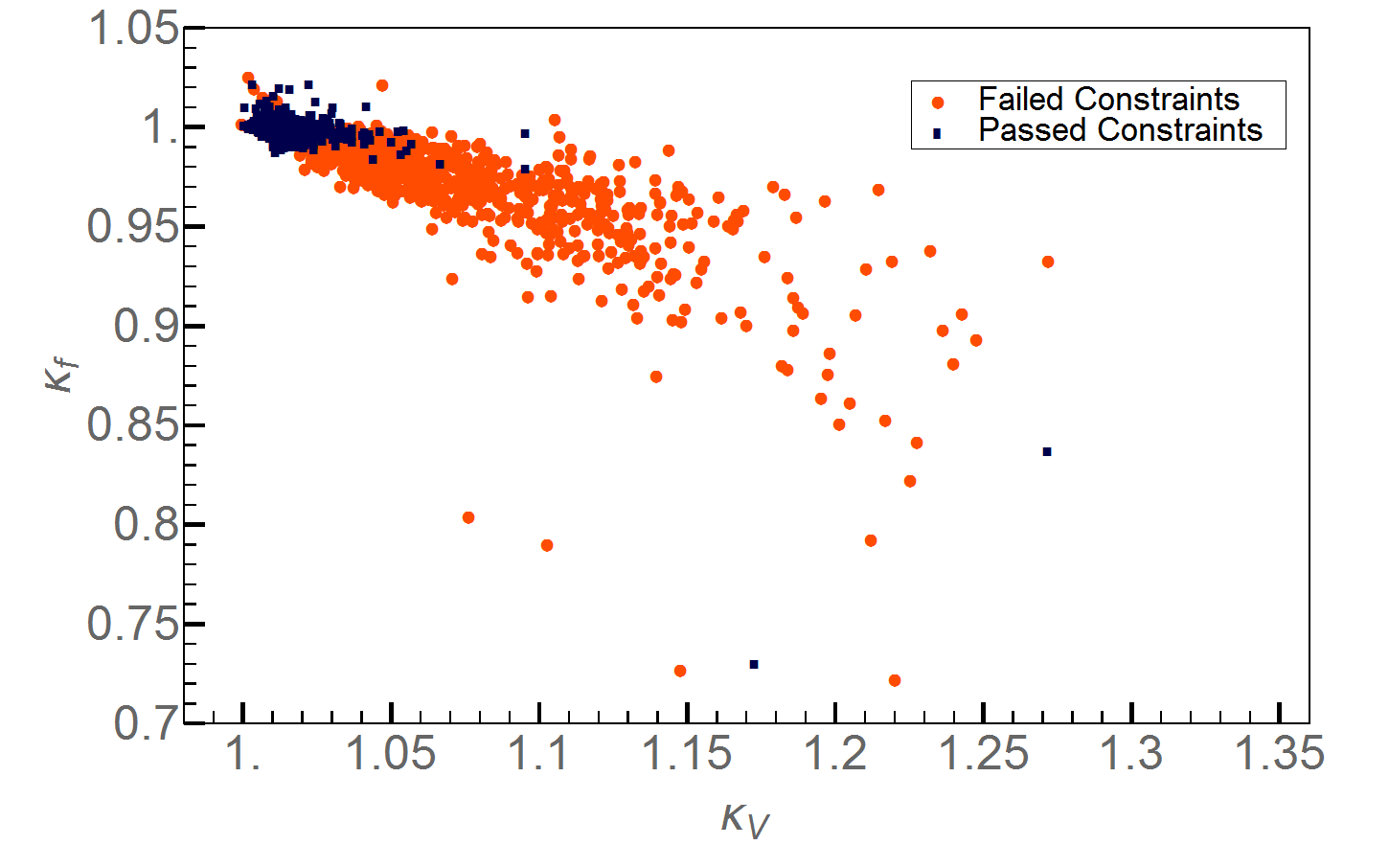}
   \caption{Couplings of $h$ to fermions ($\kappa_f^h$) and $W$ and $Z$ bosons ($\kappa_V^h$) normalized to their SM values, for only the points allowed by the dark matter direct and indirect detection constraints.  The dark blue (black) points satisfy the $\Delta \chi^2 < 4$ condition for the Higgs coupling measurements from the LHC Run 1 data~\cite{ATLAS:2016HiggsSignals}, while the red (grey) points fail this condition.}
   \label{fig:kfkv}
\end{figure}

We apply the constraints from the Higgs couplings fit only to the points that have passed 
all previous dark matter constraints. Figure~\ref{fig:kfkv} shows the points that pass (blue/black) and those that 
fail (red/gray) the $\chi^2$ constraint from Higgs couplings in the $\kappa_V^h$-$\kappa_f^h$ plane.  The Higgs coupling measurements exclude points for which $\kappa_V^h$ or $\kappa_f^h$ deviate too much from their SM value of 1.

\begin{figure}[t]
   \centering
   \includegraphics[width=0.45\textwidth]{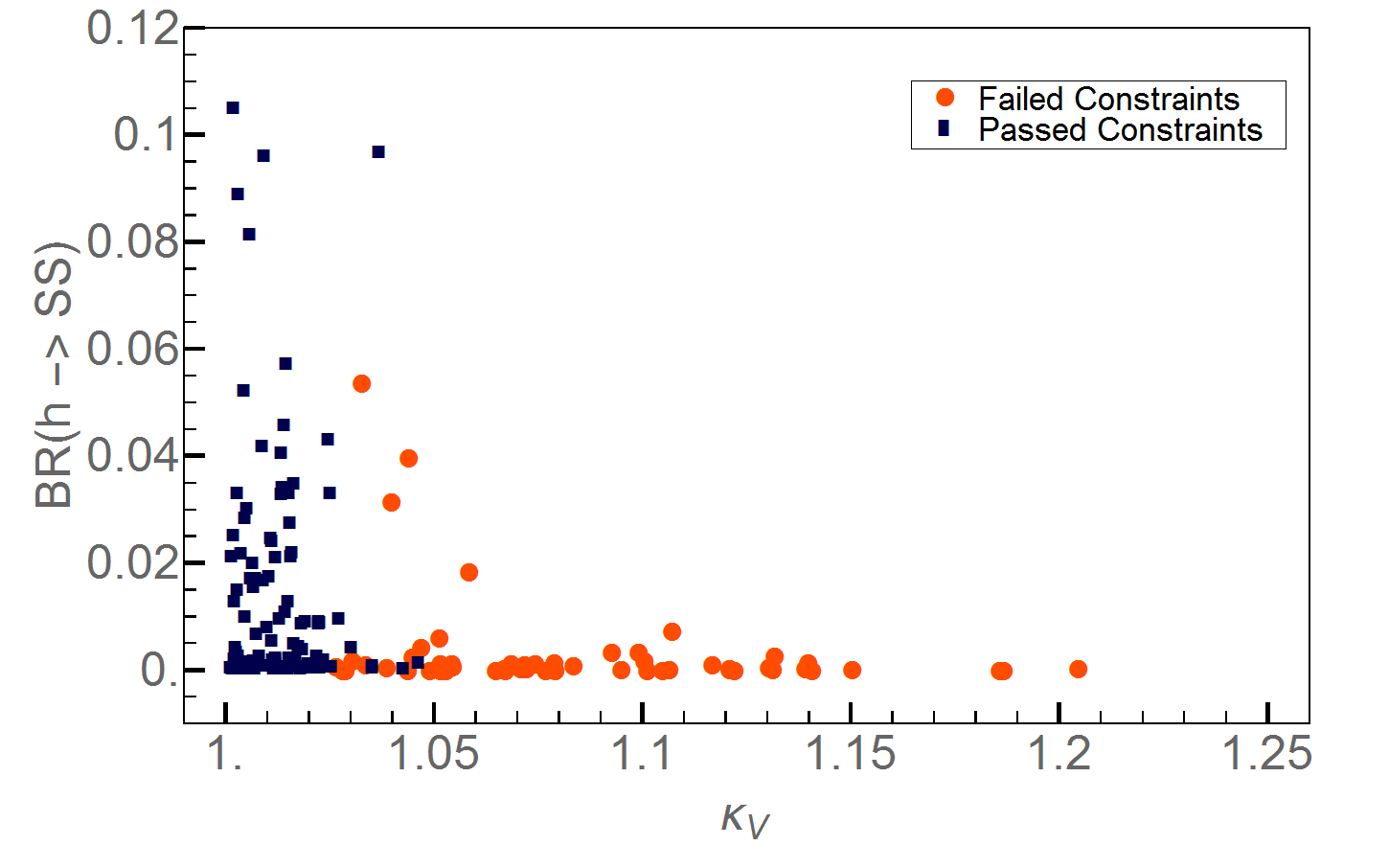}
   \caption{Branching ratio of $h \to SS$ as a function of $\kappa_V^h$, for only the points allowed by the dark matter direct and indirect detection constraints and for which $m_S < m_h/2$.  The dark blue (black) points satisfy the $\Delta \chi^2 < 4$ condition for the Higgs coupling measurements from the LHC Run 1 data~\cite{ATLAS:2016HiggsSignals}, while the red (grey) points fail this condition.}
   \label{fig:invBr}
\end{figure}

Of particular interest for Higgs phenomenology is the case where the singlet is lighter than 
half the Higgs mass. This allows the Higgs to decay to a pair of singlets which would then 
escape the detector. Figure \ref{fig:invBr} shows the branching ratio of $h\rightarrow SS$ 
as a function of $\kappa_V^h$ for the points that passed all previous constraints and have 
a singlet mass less than half the Higgs mass.

\begin{figure}[b]
   \centering
   \includegraphics[width=0.45\textwidth]{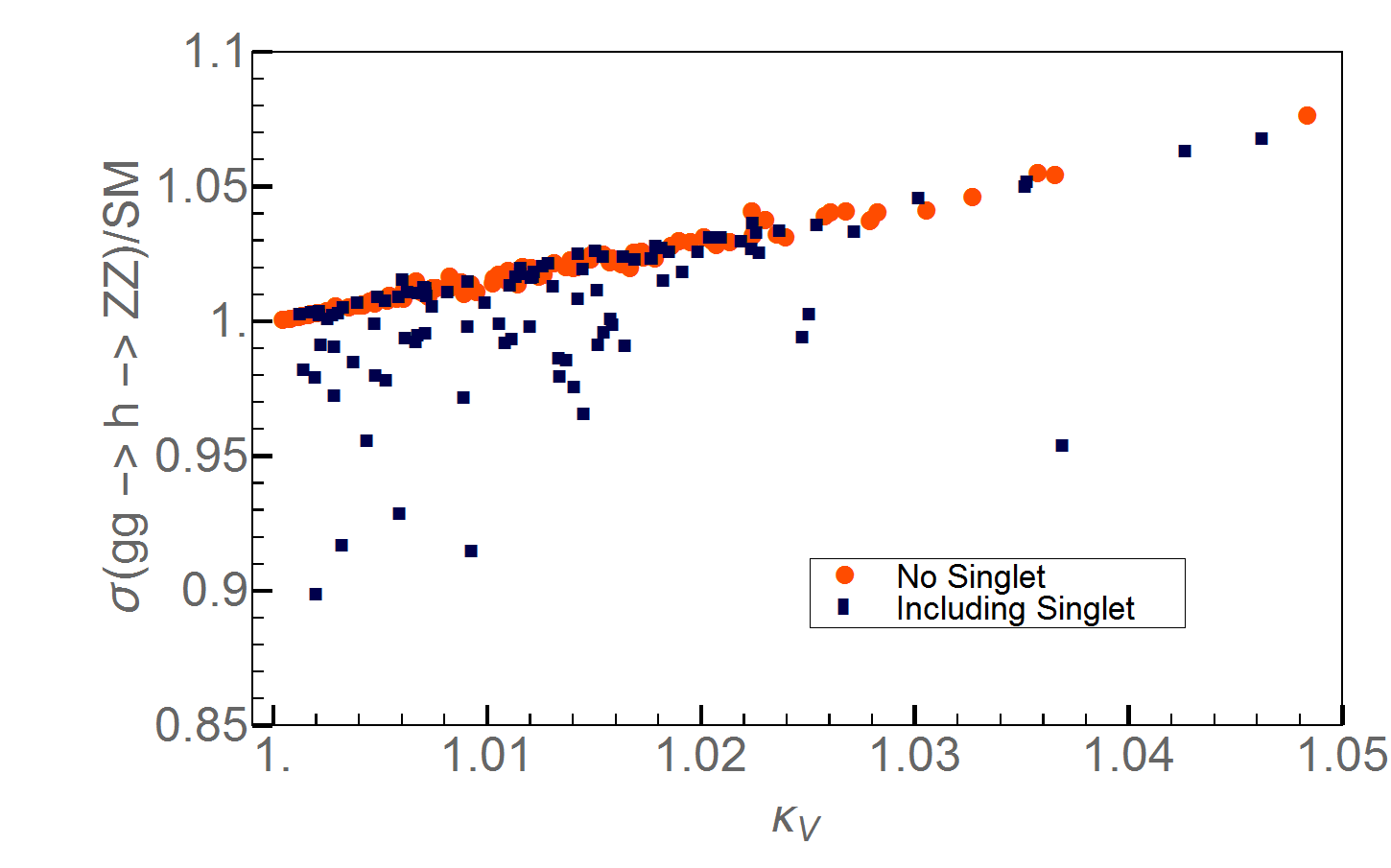}
   \caption{Prediction for the cross section $\sigma(gg \to h \to ZZ)$ normalized to its SM value as a function of $\kappa_V^f$.  The dark blue (black) points are the points that pass the dark matter direct and indirect detection constraints and satisfy the $\Delta \chi^2 < 4$ constraint from the Higgs coupling measurements.  The red (gray) points are the same parameter points but with the $h \to SS$ decays switched off (see text for details) -- these correspond to the predictions in the original GM model without the singlet scalar.}
      \label{fig:gghZZ}
\end{figure}

The only observable in the $\chi^2$ analysis that is sensitive to the total decay width of the 
Higgs boson is the $\sigma(gg\rightarrow h \rightarrow ZZ)$ cross section, because the total 
width cancels out in all the other inputs. In particular, we have:
\begin{equation}
\frac{\sigma(gg\rightarrow h \rightarrow ZZ)}{\sigma^{SM}(gg\rightarrow h \rightarrow ZZ)} = \left(\kappa_f^h \kappa_V^h\right)^2\frac{\Gamma^{SM}_{\rm tot}}{\Gamma_{\rm tot}}.
\end{equation}
This observable allows us to potentially distinguish between our model and the original GM model without the scalar singlet dark matter candidate. In Fig.~\ref{fig:gghZZ} we plot this observable versus $\kappa_V^h$ for the points that survive the $\Delta \chi^2 < 4$ constraint and for which the mass of the singlet is less than half the Higgs mass, so that $h \to SS$ is kinematically allowed.  These are the blue (black) points.  We then take the same points, set $\mu_s=\lambda_a=\lambda_b=\lambda_S=0$ while keeping the other Lagrangian parameters fixed, and remove the singlet from the theory.  These points are plotted in red (gray).  For these points the couplings $\kappa_f^h$, $\kappa_V^h$, and $\kappa_{\gamma}^h$ are the same as in the full model, but ${\rm BR}(h \to SS)$ (and its contribution to the Higgs total width) is eliminated.

As can be seen, for the original GM model without the singlet, the red points fall 
roughly along a line due to the correlation between $\kappa_V^h$ and ${\rm BR}(h \to ZZ)$ after the rest of the Higgs coupling measurements are imposed.  For the full GM model with the singlet scalar dark matter candidate, however, some of the points are scattered downward to smaller values of $\sigma(gg \to h \to ZZ)/{\rm SM}$, due to the suppression of ${\rm BR}(h \to ZZ)$ by the competing $h \to SS$ decay mode.  These are the same points for which ${\rm BR}(h \to SS)$ is visibly above zero in Fig.~\ref{fig:invBr}.  This offers a second way to potentially discriminate between the original GM model and its scalar singlet extension through an improved precision on the measurement of $\sigma(gg\rightarrow h \rightarrow ZZ)$, when $m_S < m_h/2$.

\section{Conclusions}
\label{sec:conclusions}

In this paper we studied the addition of a scalar dark matter candidate to the Georgi-Machacek model.  This provides a concrete implementation of a scenario in which the Higgs couplings to vector boson and fermion pairs can be enhanced while a new, non-SM decay mode is also present, thereby allowing an exploration of the interplay of Higgs production and decay constraints.  We showed that the dark matter candidate in this model can be made to respect all current constraints while allowing for a sizable (up to 10\%) branching ratio for the Higgs to the dark matter candidate in certain areas of parameter space.

The model consists of the Georgi-Machacek model with the addition of a real 
singlet which has a $Z_2$ symmetry to make it stable. We first studied
the theoretical constraints on the new parameters by imposing tree-level 
unitarity in $2\rightarrow 2$ scalar scattering amplitudes, requiring that the potential is bounded from below, and requiring that deeper custodial $SU(2)$-violating minima are absent. 
We found that we could translate all the constraints from the original GM model to our 
extended model and simply add a few new constraints on the new Lagrangian parameters. 

We performed a numerical scan over the Lagrangian parameters, imposing the theoretical 
constraints and requiring that the singlet scalar accounts for all of the dark matter in the universe through thermal freeze-out.  We identified the parameter regions that satisfy the constraints from dark matter direct-detection searches as well as the indirect constraints from gamma ray measurements of dwarf spheroidal galaxies.  This constrained the dark matter mass to be either near the Higgs pole for resonant 
annihilation (57-62 GeV) or mostly above about 120 GeV. We also saw that for parameter values 
where the dark matter mass was near the Higgs pole we could attain a sizable branching 
ratio for $h\rightarrow SS$; however, after imposing the dark matter constraints, the current limit on the Higgs invisible decays does not further constrain the model.

We finally studied the constraints from the LHC Run 1 Higgs coupling measurements.  While these measurements further constrain the parameter space, they do so mostly by constraining the $hf \bar f$, $h VV$, and $h \gamma\gamma$ couplings.  The Higgs coupling measurements are not yet precise enough to be sensitive to the modification of signal rates by the presence of the $h \to SS$ decay mode, so that the constraints from Higgs measurements are so far the same as they would be in the original GM model without the singlet scalar.

The allowed region of parameter space that we identified can be further probed in the future
by the next generation of dark matter direct detection experiments, as well as improved precision on the invisible Higgs decay width and Higgs coupling measurements.

\begin{acknowledgments}
We thank Travis Martin and Jonathan Kozaczuk for helpful conversations.
This work was supported by the Natural Sciences and Engineering Research Council of Canada.  
H.E.L.~also acknowledges support from the grant H2020-MSCA-RISE-2014 no.~645722 (NonMinimalHiggs).
\end{acknowledgments}

\appendix
\section{Feynman rules for couplings involving $S$}
\subsection{Triple scalar couplings}
The Feynman rules for couplings to $S$ are given by $-ig_{SSh_i}$ with all particles incoming and the couplings defined as follows:
\begin{align}
g_{SSh}&=-4\left(\lambda_a s_\alpha v_\phi+\sqrt{3}\lambda_b c_\alpha v_\chi\right),\notag\\
g_{SSH}&=-4\left(\lambda_a c_\alpha v_\phi+\sqrt{3}\lambda_b s_\alpha v_\chi\right),
\end{align}
where we use the  notation $s_\alpha\equiv\sin \alpha$ and $c_\alpha\equiv\cos \alpha$.

\subsection{Quartic scalar couplings}
The Feynman rules for couplings to $S$ are given by $-ig_{SSs_1s_2}$ with all particles incoming and the couplings defined as follows:
\begin{align}
g_{SShh}&=-4\left(\lambda_a c^2_\alpha+\lambda_b s^2_\alpha \right), \notag \\
g_{SSHH}&=-4\left(\lambda_a s^2_\alpha+\lambda_b c^2_\alpha \right), \notag \\
g_{SShH}&=-4s_\alpha c_\alpha\left(\lambda_a - \lambda_b  \right), \notag \\
g_{SSG^0G^0}&=g_{SSG^+G^{+*}}=-4\left(\lambda_a c^2_H+\lambda_b s^2_H \right), \notag \\
g_{SSH_3^0H_3^0}&=g_{SSH_3^+H_3^{+*}}=-4\left(\lambda_a s^2_H+\lambda_b c^2_H \right), \notag \\
g_{SSG^0H_3^0}&=g_{SSG^+H_3^{+*}}=g_{SSH_3^+G^{+*}}=-4s_H c_H\left(\lambda_b -\lambda_a \right), \notag \\
g_{SSH_5^0H_5^0}&=g_{SSH_5^+H_5^{+*}}=g_{SSH_5^{++}H_5^{++*}}=-4\lambda_b,
\end{align}
where we use the notation $s_H\equiv\sin \theta_H$ and $c_H\equiv\cos \theta_H$, and $G^0$ and $G^{\pm}$ are the Goldstone bosons.

All other Feynman rules are identical to those in the original GM model and can be found in Appendix A of Ref. \cite{Hartling:2014zca}


\end{document}